\documentclass[manuscript]{acmart}

\usepackage{hyperref}
\hypersetup{
 unicode=true,
 pdfstartview={FitH},
 colorlinks=true,
 citecolor=blue,
 linkcolor=blue,
 urlcolor=blue
}

\usepackage[linesnumbered,algoruled]{algorithm2e}
\usepackage{algpseudocode}

\usepackage{caption}
\usepackage{subcaption}
\usepackage{makecell}

\usepackage{graphicx}
\usepackage{multirow}
\usepackage{booktabs}
\usepackage[comments]{annotations}
\newcommand{\lc}[1]{\mynote{LC}{#1}{blue!20!magenta}}

\newcommand{\nop}[1]{}
\newcommand{\revision}[1]{{\color{red}{#1}}}
\newcommand{\Com}[1]{}

\usepackage{amsthm}
\usepackage{listings}

\usepackage{cleveref}
\crefname{algocf}{alg.}{algs.}
\Crefname{algocf}{Algorithm}{Algorithms}
\SetKwRepeat{Do}{do}{while}

\usepackage{enumitem}
\usepackage{acronym}
\acrodef{IP}[IP]{intellectual property block}
\acrodef{SoC}[SoC]{System-on-Chip}
\acrodef{IC}[IC]{integrated circuit}
\acrodef{eFPGA}[eFPGA]{embedded field programmable gate array}
\acrodef{FPGA}[FPGA]{field programmable gate array}
\acrodef{RTL}[RTL]{register transfer level}
\acrodef{CLB}[CLB]{configurable logic block}
\acrodef{LUT}[LUT]{look-up table}
\acrodef{HLS}[HLS]{high-level synthesis}
\acrodef{EDA}[EDA]{electronic design automation}
\acrodef{FF}[FF]{flip-flop}
\acrodef{DIP}[DIP]{distinguishing input pattern}
\acrodef{PPA}[PPA]{power, performance, and area}
\acrodef{BLE}[BLE]{basic logic element}
\acrodef{CB}[CB]{connection block}
\acrodef{SB}[SB]{switch block}
\usepackage{booktabs}
\usepackage{tabularx}
\newcolumntype{L}[1]{>{\raggedright\let\newline\\\arraybackslash\hspace{0pt}}m{#1}}
\newcolumntype{C}[1]{>{\centering\let\newline\\\arraybackslash\hspace{0pt}}m{#1}}
\newcolumntype{R}[1]{>{\raggedleft\let\newline\\\arraybackslash\hspace{0pt}}m{#1}}

\usepackage{pifont}
\input{preamble}

\setcopyright{acmlicensed}
\copyrightyear{2025}
\acmYear{2025}
\acmDOI{XXXXXXX.XXXXXXX}

\setcopyright{acmlicensed}
\acmJournal{TODAES}
\acmYear{2025} \acmVolume{1} \acmNumber{1} \acmArticle{1} \acmMonth{1}\acmDOI{10.1145/3737287}

\begin{document}

\title{ARIANNA: An Automatic Design Flow for Fabric Customization and eFPGA Redaction}

\author{Luca Collini}
\email{lc4976@nyu.edu}
\author{Jitendra Bhandari}
\email{jb7410@nyu.edu}
\affiliation{%
  \institution{New York University}
  \city{New York}
  \state{New York}
  \country{USA}
}
\author{Chiara Muscari Tomajoli}
\email{chiara.muscari@mail.polimi.it}
\affiliation{%
  \institution{Politecnico di Milano}
  \city{Milano}
  \country{Italy}
}
\author{Abdul~Khader~Thalakkattu~Moosa}
\email{at4856@nyu.edu}
\affiliation{%
  \institution{New York University}
  \city{New York}
  \state{New York}
  \country{USA}
}
\author{Benjamin Tan}
\email{benjamin.tan1@ucalgary.ca}
\affiliation{%
  \institution{University of Calgary}
  \city{Calgary}
  \state{Alberta}
  \country{Canada}
}
\author{Xifan~Tang}
\email{xifan@rapid-flex.com}
\affiliation{%
  \institution{Rapid Flex}
  \city{San Jose}
  \state{California}
  \country{USA}
}
\author{Pierre-Emmanuel~Gaillardon}
\email{pegaillardon@cade.utah.edu}
\affiliation{%
  \institution{University of Utah}
  \city{Salt Lake City}
  \state{Utah}
  \country{USA}
}

\author{Ramesh Karri}
\email{rkarri@nyu.edu}
\affiliation{%
  \institution{New York University}
  \city{New York}
  \state{New York}
  \country{USA}
}

\author{Christian Pilato}
\email{christian.pilato@polimi.it}
\affiliation{%
  \institution{Politecnico di Milano}
  \city{Milano}
  \country{Italy}
}

\renewcommand{\shortauthors}{Collini et al.}


\begin{abstract}
In the modern global Integrated Circuit (IC) supply chain, protecting intellectual property (IP) is a complex challenge, and balancing IP loss risk and added cost for theft countermeasures is hard to achieve. Using embedded configurable logic allows designers to completely hide the functionality of selected design portions from parties that do not have access to the configuration string (bitstream). However, the design space of redacted solutions is huge, with trade-offs between the portions selected for redaction and the configuration of the configurable embedded logic. We propose ARIANNA, a complete flow that aids the designer in all the stages, from selecting the logic to be hidden to tailoring the bespoke fabrics for the configurable logic used to hide it. We present a security evaluation of the considered fabrics and introduce two heuristics for the novel bespoke fabric flow. We evaluate the heuristics against an exhaustive approach. We also evaluate the complete flow using a selection of benchmarks. Results show that using ARIANNA to customize the redaction fabrics yields up to 3.3$\times$ lower overheads and 4$\times$ higher eFPGA fabric utilization than a one-fits-all fabric as proposed in prior works.
\end{abstract}

\maketitle


\section{Introduction}

Securing hardware Intellectual Property (IP) is a crucial concern during the design and production of an Integrated Circuit (IC)~\cite{9310331}. With the multi-billion dollar investments required for cutting-edge manufacturing plants, many design houses are forced to outsource IC fabrication to external foundries. 
This situation has enormous security implications as an unscrupulous employee can steal the IC design to reverse engineer it and make illegal copies~\cite{DBLP:journals/todaes/ShamsiLPFPJ19}. Design houses can use several techniques, like watermarking, split manufacturing, and logic locking, to safeguard their blueprints~\cite{9310331}. However, these strategies are not foolproof: watermarking is a \textit{passive} method aimed to identify an IP theft after it has occurred~\cite{survey_watermarking}; split manufacturing requires high manufacturing expertise~\cite{9216128}; and logic locking is susceptible to a wide array of security breaches~\cite{DBLP:journals/todaes/ShamsiLPFPJ19,DBLP:journals/corr/abs-2006-06806}, especially when the attacker can access a working chip (called \textit{oracle}).


\textbf{FPGA redaction} is an innovative, promising method to counter reverse engineering attempts. This approach proposes substituting critical portions of the design with specially designed reconfigurable blocks, also known as \textit{embedded FPGAs} (eFPGAs). The dual objectives of this technique are: (1) during fabrication, masking the ultimate functionality of the reconfigurable block behind the ambiguity of its reconfigurable nature; (2) during execution, implementing the original functionality by loading the correct bitstream.

%

\setlength{\textfloatsep}{8pt}
\begin{figure}[t]
\centering
\includegraphics[width=0.65\columnwidth]{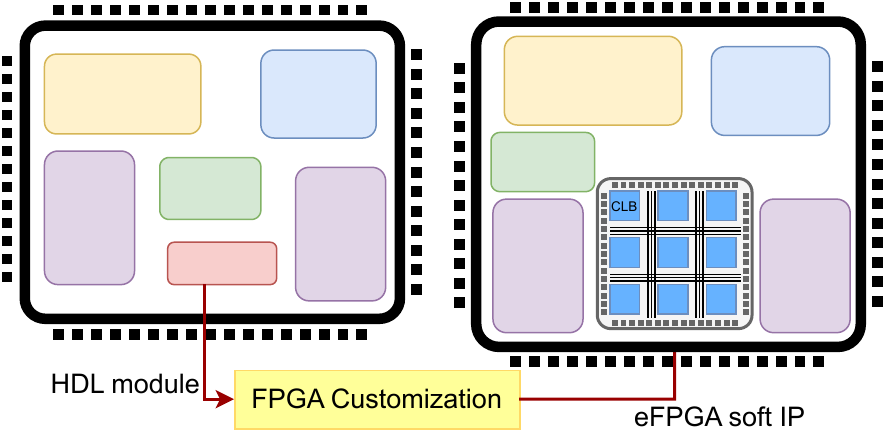}
\vspace{-8pt}\caption{FPGA redaction flow. Critical modules are replaced with custom eFPGA implementations.}\label{fig:efpga_flow}
\end{figure}

\autoref{fig:efpga_flow} shows an example where a module is replaced by a custom eFPGA fabric. Within this fabric, each unit represents a Configurable Logic Block (CLB). Cutting-edge tools for FPGA specialization (e.g., OpenFPGA~\cite{9098028} and FABulous~\cite{DBLP:conf/fpga/KochDHYA21}) allow engineers to map an HDL module into a soft eFPGA IP. This resultant IP can be seamlessly combined and synthesized with the chip's remaining components. The resilience of this FPGA-obfuscation method against SAT attacks is attributed to the need to decode a large set of ``key bits'' —essentially the complete configuration stream of the eFPGA— through the intricate interactions between input and output within the eFPGA structure~\cite{mohan_hardware_2021,our_iccad_21}. Additionally, these tailor-made eFPGAs present a lower overhead than standard, commercially available options~\cite{our_iccad_21,9369856}.

%


The process of FPGA redaction involves multiple stages for designers. Initially, they must determine which modules are optimal for redaction, considering both security implications and design perspectives. Following that, the task is to design and assimilate the tailored eFPGA fabric into the system. These challenges are intricately linked and frequently vary based on the specific application. Given the complexities, designers typically address these issues manually, which can result in suboptimal solutions~\cite{9369856,Chen_dac_2020}.

Previous work~\cite{alice} focused on the EDA problem of \textit{partitioning RTL modules} between eFPGA and ASIC and \textit{creating the proper eFPGA fabrics} to implement the redacted modules. While the malicious foundry can retrieve modules implemented in ASIC, the flexibility of eFPGAs protects the redacted modules. In prior work~\cite{alice}, the eFPGA fabric configuration was kept fixed for all redacted designs, resulting in solutions that required more resources than necessary. The parameters for eFPGA fabrics are numerous, and the design space is vast, making its exploration a hard problem. In this work, we propose \textbf{ARIANNA} (\uline{A}utomatic eFPGA \uline{R}edact\uline{I}on with f\uline{A}bric co\uline{N}figuratio\uline{N} \uline{A}nd module clustering), a \textbf{complete flow to identify the modules to be redacted, optimize the eFPGA fabric, and generate the corresponding chip design augmented with soft eFPGAs}. 
ARIANNA performs a progressive refinement of the solution by identifying candidate modules for redaction and clustering them to enable the creation of larger eFPGAs, whose fabric configurations are explored and characterized to select the best final implementation that minimizes the hardware overhead without sacrificing security.
ARIANNA is built on top of the ALICE framework~\cite{alice}. While ALICE focuses on identifying the modules for redaction and generating their corresponding soft eFPGAs with a pre-defined configuration, ARIANNA extends the ALICE approach by \textbf{defining the ultimate set of secure fabrics} and \textbf{fine-tuning the design-specific eFPGA fabric} to minimize the hardware cost.
Our contributions can be summarized as follows:
\begin{itemize}[leftmargin=1.5em]
\item We build upon the ALICE framework to include heuristics for \textbf{eFPGA fabric parameter optimization} for hardware efficiency, obtaining a more holistic and fine-grained approach to hardware IP protection (\autoref{sec:heuristics}).
\item We present a \textbf{security evaluation} of eFPGA fabric as a pre-step to identify the configurations to explore with the novel fabric tailoring heuristics (\autoref{sec:sec_identification}, \autoref{sec:exp_sec}).
\item We validate the proposed heuristics through extensive benchmarks against an exhaustive approach, showing their efficacy and efficiency (\autoref{sec:exh}). 
\item We validate the complete framework on a set of benchmarks to show the efficacy of our \textbf{fine-tuned eFPGA fabrics} in \textbf{reducing the overheads} of hardware IP protection (\autoref{sec:eval}).
\end{itemize}
This work effectively bridges the gap between module selection and eFPGA fabric optimization by focusing on \textbf{tailoring the eFPGA fabric}. Designers now have a more comprehensive toolset to consider functional characteristics, structural attributes, and eFPGA parameters, facilitating a more robust and efficient redaction process.

The rest of the paper is organized as follows. \autoref{sec:fundamentals} introduces the background notions on custom eFPGA design flows and related architectures. In \autoref{sec:related_work}, we provide an overview of related works on IP protection with embedded FPGAs and attacks against them, while the threat model is defined in \autoref{sec:threat_model}. In \autoref{sec:alice}, we present the proposed framework for eFPGA redaction, while in \autoref{sec:results} we evaluate the proposed heuristics for bespoke fabrics and the complete framework. In \autoref{sec:conclusion}, we make our final comments and takeaways on our work.

\section{Fundamentals on Embedded FPGAs \label{sec:fundamentals}}
This section briefly overviews (embedded) FPGAs, emphasizing significant elements such as architectural alternatives and open-source toolchains that aid in versatile hardware development techniques. These facets are critical for creating tailored eFPGA solutions, especially for IP redaction. For a more in-depth exploration of FPGA architectures, we recommend consulting the comprehensive study by Boutros and Betz~\cite{FPGA_arch}.

\begin{figure}[htb]
\centering
\hspace{-20pt}
\includegraphics[width=1\columnwidth]{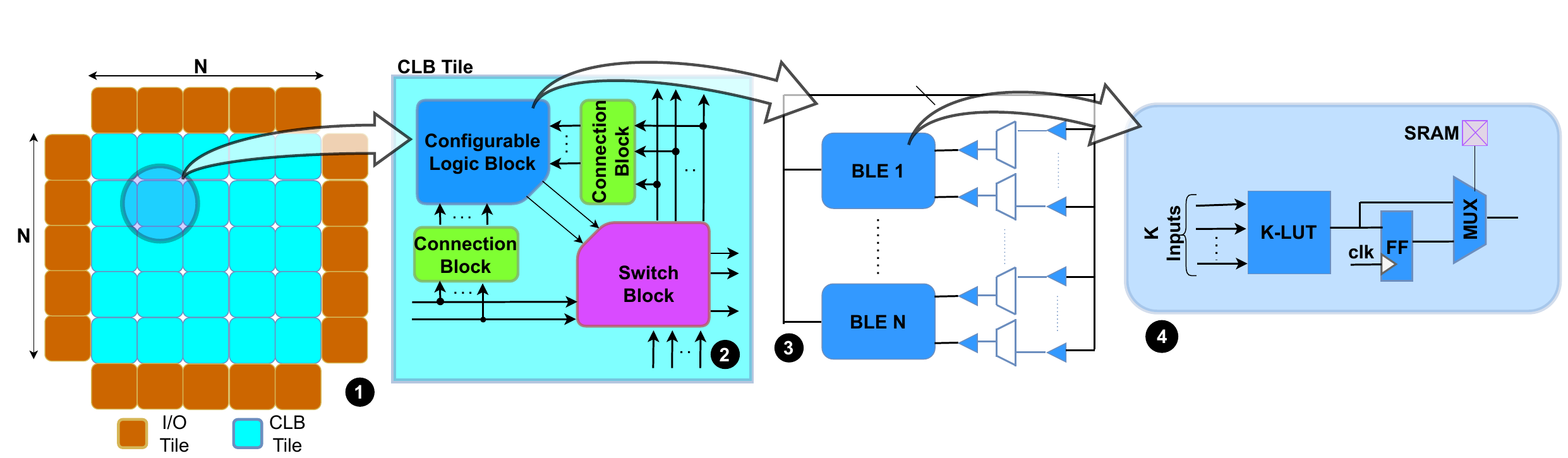}
\caption{A general FPGA architecture and its constituent parts. }
\label{fig:fpga}
\end{figure}

\subsection{FPGA Architectures\label{sec:fpga_arch}}

FPGAs are flexible architectures capable of being reprogrammed ``on-site'' (in the field) to manifest a particular digital design. Modern FPGAs utilize a tile-based architecture, consisting of recurrent tiles and a ``sea'' of routing resources, as depicted in \autoref{fig:fpga}~(\ding{202}). 
An \textit{N$\times$N} architecture means there are \textit{N} tiles distributed in horizontal and vertical directions, respectively. 
 \ac{CLB} tiles are predominant in an FPGA and implement logic functions (both combinational and sequential). Modern FPGAs can also include specialized tiles, such as block RAM (BRAM) or digital signal processing (DSP) tiles, for storing data on-chip and performing efficient arithmetic operations. A heterogeneous tile-based FPGA allows designers to satisfy design needs while managing the aspects of \ac{PPA} within the architecture. Tile-based architectures present a more favorable balance between programmability and efficiency relative to other options~\cite{FPGA_arch}; this structure enables designers to concentrate individually on the challenges of routing and connecting signals inside a tile and the issue of ``globally'' interlinking tiles. Consequently, engineers can prioritize the optimization of a tile's layout, reducing the time spent on the placement and routing of tiles.

 \autoref{fig:fpga}~(\ding{203}) details a \ac{CLB} tile. It contains a \ac{CLB} and different blocks for setting the connection between signals within and outside the tile. Refer to \autoref{fig:fpga}~(\ding{204}) for an in-depth view of the \ac{CLB} structure. It comprises $N$ \acp{BLE}, interconnected via a local routing system. Each \ac{BLE} is the basic unit for logic operation and encompasses a \ac{LUT}, a \ac{FF}, and a 2-input multiplexer, as depicted in \autoref{fig:fpga} (\ding{205}). A single-output Boolean function with $K$ inputs can be mapped onto an LUT with the same number of inputs.



By configuring a 2-input multiplexer, a \ac{BLE} can operate in combinational or sequential mode. 
To route interconnect \ac{CLB} inputs and \ac{BLE} inputs and outputs, the local routing architecture, typically implemented as a crossbar, includes a set of programmable multiplexers. The local routing guarantees that \acp{BLE} can be fully connected to each other and to every \ac{CLB} input pin. 
To optimize the hardware cost, in this work, we focus on the structure of the \ac{CLB}.
The capability of a \ac{CLB} is defined by these factors:
(1) the input dimension of the LUTs, denoted as $K$;
(2) the quantity of \ac{BLE} within a \ac{CLB}, represented as $N$;
and (3) the total input count for the \ac{CLB}, labeled $I$.
The selection of these parameters is influenced by the balance between logic capability and its effects on size, timing, and power consumption.
To have better resource utilization in a \ac{CLB}, for any LUT size, $I = \frac{K(N+1)}{2}$ has been shown to give good \ac{PPA}~\cite{LUT}. 
For these reasons, we explore $N$ and $K$ as the two parameters of the eFPGA fabric configuration in this work. 

The FPGA is programmed using a \textit{bitstream}, in which each bit dictates certain fabric components, like routing setups and \ac{LUT} data. The bitstream can be uploaded using frame-based methods~\cite{koch_fabulous_2021} or scan-chain-based methods~\cite{mohan_top-down_2021,tang_openfpga_2019}. For our research, we concentrate on the scan-chain based bitstream loading. In this approach, the complete bitstream is input sequentially, with each clock cycle loading one bit, utilizing a specific clock designated for this task.

\subsection{Custom eFPGA Design Flow}
Reconfigurable devices can implement any specific function by loading the proper configuration bitstream. In the case of hardware security, this post-manufacturing adaptability is crucial for safeguarding hardware \ac{IP}s. Designers can embed the FPGAs into ASIC designs as ready-made blocks, with only the end-user handling their configuration. As a result, the function executed by the FPGA remains undisclosed to the manufacturing facility that cannot access the correct configuration bitstream.

\begin{figure}
\centering
\includegraphics[width=0.65\columnwidth]{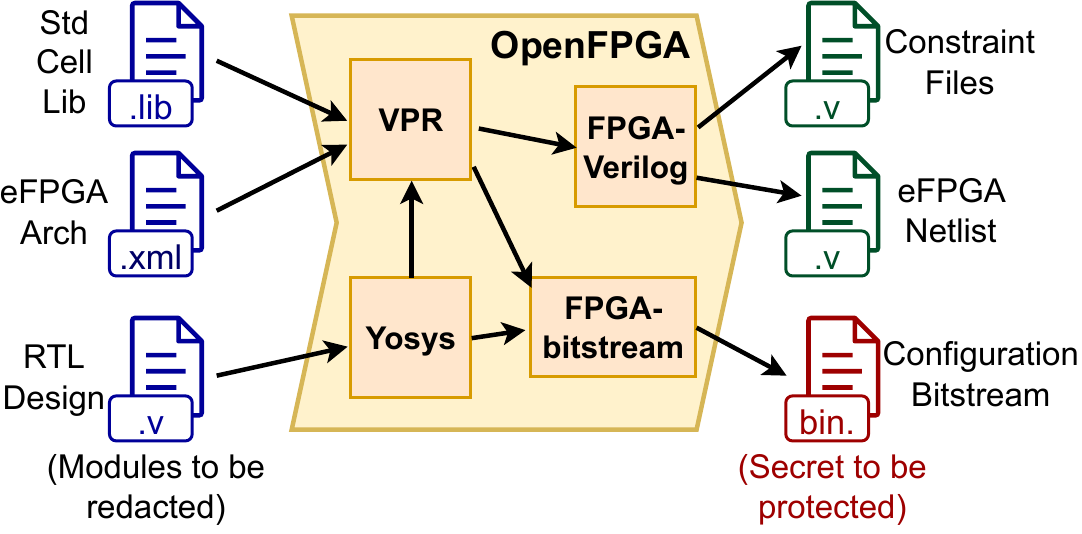}
\vspace{-6pt}\caption{ARIANNA leverages a state-of-the-art eFPGA design flow based on OpenFPGA~\cite{9098028}. The eFPGA netlist is integrated with the rest of the chip, while the configuration bitstream is kept secret.}\label{fig:fpga_flow}
\end{figure}

Recently, open-source eFPGA prototyping tools are becoming increasingly popular~\cite{tang_openfpga_2020, koch_fabulous_2021, ALi_FPGA_2021}. These platforms facilitate the automated adaptation of FPGA structures, specially designed for distinct modules, encompassing the entire Verilog-to-bitstream process. For instance, \Cref{fig:fpga_flow} illustrates the customization process based on OpenFPGA, suitable for eFPGA redaction~\cite{our_iccad_21}.
OpenFPGA utilizes an XML-defined fabric parameter to generate the associated eFPGA IP ready for fabrication~\cite{9098028, our_iccad_21, JLuu_FPGA_2011, XTang_tvlsi_2018}. The modules set for modification influence the eFPGA customization\footnote{For a complete overview of the OpenFPGA flow, please see the tool \href{https://openfpga.readthedocs.io/en/master/}{documentation}.}. Leveraging open-source platforms provides designers with enhanced flexibility, allowing adjustments to many parameters, as highlighted in prior work~\cite{bhandari2021fabrics}. This empowers users to devise architectures optimally tailored to the intended design. When integrated closely with processors on a unified chip, these structures can function as adaptable accelerators or co-processors \cite{mohan_hardware_2021, 9369856, intel_xeon_fpga, PSchiavone_tvlsi_2021}. This integration can enhance a \ac{SoC} peak performance by up to 3.4$\times$ while reducing power consumption by approximately 2.9$\times$.

Our study delves into FPGA structures characterized by varying parameters $K$ and $N$ within \ac{CLB}s to obtain the most cost-efficient structure for redacting the given design. However, we can accommodate any fabric setup or even explore other parameters similarly, as our primary emphasis is on their application for redaction rather than their creation or security assessment. 

\section{Motivation and Related Work}\label{sec:related_work}

The protection of hardware IPs has become a major focus in recent years.
Researchers proposed many methods, especially at low levels of abstraction (i.e., on gate-level netlists or physical designs, or directly during fabrication~\cite{8050883,eASIC}. For example, logic locking assumes the attacker cannot retrieve the correct functionality thanks to the protection of a ``secret'', the locking key~\cite{8852678}. Despite many advances~\cite{8203496, llfrontiers, llanalysis}, SAT attacks~\cite{7140252} and machine learning attacks~\cite{omla,unsail,neuro,rtlml} can be used to identify the I/O relationships and retrieve key bits when an activated chip is available, challenging the effectiveness of logic locking~\cite{DBLP:journals/todaes/ShamsiLPFPJ19,DBLP:journals/corr/abs-2006-06806}.


FPGA redaction is a recent technique that aims to implement selected modules with soft or hard eFPGAs that are included in the design. The key idea is that (1) attackers in the foundry have no access to the bitstream configuration that can implement any possible functionality, while (2) end-user attackers that have access to an activated chip cannot retrieve the correct bitstream. In this case, the ``secret'' corresponds to the configuration bitstream. While eFPGA redaction is considered more secure than logic locking, the design of FPGA-redacted ICs is complex, especially in the module partitioning between eFPGA and ASIC. Moreover, eFPGA redaction comes with higher overhead costs than logic locking techniques. This work proposes a complete flow that helps designers find a feasible module combination for eFPGA redaction, minimizing the overheads.

While recent studies focused on VLSI challenges of eFPGA integration~\cite{Mohan_fpga_2021}, selecting the modules to be redacted is still a manual effort or requires at least a reference design. In the former case, designers have to identify the modules to be protected, for example, because they are part of the core business~\cite{mohan_hardware_2021}. Designers may want to use FPGA redaction to protect the results of selected outputs with FPGA redaction without knowing the critical components.
In the latter case, two or more designs are compared with each other to identify common parts (which are assumed to be common to many other designs) and different parts (which are the unique parts of the given design)~\cite{Chen_dac_2020}. However, designers may not have an alternative version of the same design to be compared with.

SheLL~\cite{shell} is a framework proposed to reduce redaction overheads. In SheLL authors point out how the OpenFPGA framework can lead to suboptimal solutions with unutilized tiles and propose the use of the Fabulous\cite{koch_fabulous_2021} for redaction, together with a custom flow for mapping the logic to be redacted onto the LUTs.
In this work, we explore the OpenFPGA fabric customization to tailor the eFPGA architecture to the redacted modules to save overheads.
In reference~\cite{investigatingFabrics}, Sathe et al. highlight how the fabric choice for eFPGA redaction can drastically affect overheads without impacting security. 

In reference~\cite{rezaei}, Rezaei et al. proposed two attacks to break eFPGA redaction. The DIP exclusion attack is based on the idea of excluding input patterns that lead sequential SAT attacks like CycSAT~\cite{cyc_sat} to get stuck until the attack runs successfully. Brake \& Unroll works by first breaking the simple combinational cycles and then unrolling hard cycles. The results show that the attacks are successful only for the simpler eFPGA fabric configurations In reference~\cite{robustness}, Karmakar et al. propose variations of SAT attacks targeted to break eFPGA redaction. They validate their attack on the HeLLO CTF 2022 benchmarks, succeeding on small and medium benchmarks. These attacks all rely on the assumption that the attacker will gain full scan-chain access to break down the circuit into combinational logic cones.

FuncTeller~\cite{functeller} is an attack that aims to recover the redacted logic's functionality. The recovered functionality can then be synthesized for the eFPGA fabric under attack to obtain a bitstream. The attack aims at finding minterms for the ON-sets by performing a smart exploration of primary implicants. While results show that the success of FuncTeller does not depend on the eFPGA fabric configuration, the retrieved functionality is only approximate. FuncTeller also requires full scan chain access to have a black box view of all the combinational parts of the design. MANTIS~\cite{mantis} is a machine-learning-based attack on FPGA redaction that allows the retrieval of approximated bitstreams (10-20\% error rates). It does not require full scan chain access, but it still requires scan chain access to the eFPGA I/Os. 

Recent studies on the security of FPGA redaction show that the resilience to SAT attacks is correlated more with the eFPGA fabric configuration and its utilization rather than the implemented module(s)~\cite{our_iccad_21,mohan_hardware_2021,bhandari2021fabrics}. Moreover, from the before-mentioned attack studies, it emerges that it is possible to identify a subset of eFPGA fabrics that have shown to be resilient from reverse engineering attacks.


\section{Threat Model}\label{sec:threat_model}
The different attacks against eFPGA redaction discussed in \Cref{sec:related_work} have in common the following aspects of the threat model: oracle access with full (limited for MANITS~\cite{mantis}) scan-chain access and isolation of the eFPGA fabric. Moreover, for all attacks but FuncTeller and MANTIS, a subset of secure fabrics (complex enough that SAT-based attacks are unfeasible) is identifiable. 
We assume that designers looking to protect their IPs will also protect the scan chain either by blowing fuses after testing or by using a secure scan chain protection technique such as~\cite{disorc}. Adopting such a countermeasure will make all the attacks mentioned in \Cref{sec:related_work} unfeasible.
Our work aims to explore the design space of secure solutions (those that stand against state-of-the-art attacks) to identify good solutions in terms of I/O and CLB utilization and area/power overheads. 
Secure fabric parameters can be identified by looking at the state of the art or by performing an analysis beforehand, as we present in \autoref{sec:sec_identification}.


\section{ARIANNA Design Flow for eFPGA Redaction\label{sec:alice}} 
Our proposed framework aims to aid the partitioning of RTL designs for eFPGA redaction by exploring module cluster configurations and tailoring the best-fitting fabric for each cluster. 
\begin{figure}
\centering
\includegraphics[width=\columnwidth]{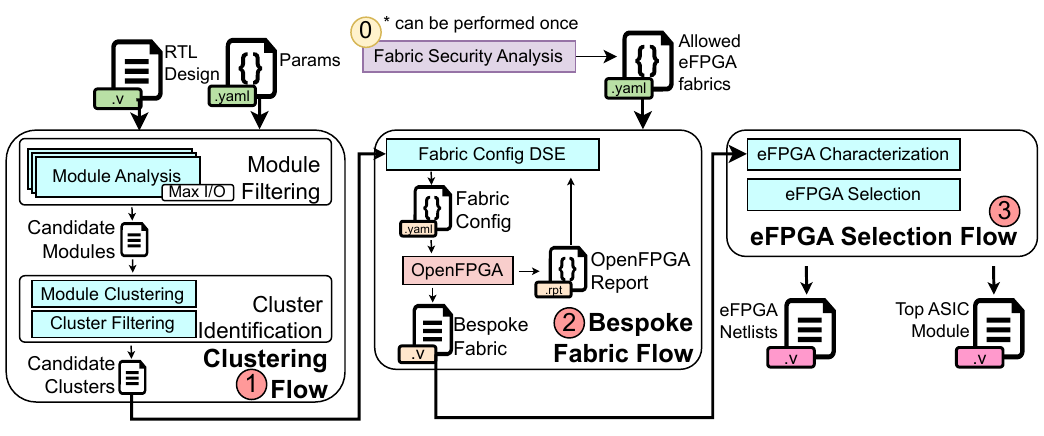}
\vspace{-4pt}\caption{ARIANNA flow for automatic eFPGA-based redaction.}\label{fig:flow}
\end{figure}
\autoref{fig:flow} shows the proposed redaction flow. It starts from a behavioral Verilog\footnote{Limitations are only due to the HDL parser we use. Supporting another HDL language (e.g., VHDL) only requires the proper parser.} RTL. A YAML configuration file provides parameters such as the file containing available eFPGA fabric configurations, the number of eFPGAs to use for redaction, and the maximum number of I/O pins for each eFPGA. The maximum number of I/O can be used to bind the module I/O to the maximum I/O size of the largest admissible eFPGA fabric. For instance, a 4$\times$4 fabric configuration has no more than 64 I/O pins~\cite{our_iccad_21,mohan_hardware_2021}. 
We support one or more eFPGAs with heterogeneous fabric configurations. 
Our ARIANNA flow is composed of three main phases:
\begin{enumerate} [start=0]
    \item \textbf{Secure Fabric Identification}: This pre-step needs to be performed offline once to identify a set of secure fabric parameters on which to perform design space exploration. Alternatively, one can look at the state-of-the-art to identify these parameters.
    \item \textbf{Clustering Flow}: This phase first identifies candidate redaction modules by filtering out modules that do not meet I/O constraints or do not affect signals of interest. It then creates clusters of candidate modules that can fit into the same eFPGA, which we refer to as \textit{candidate module clusters}.

    \item \textbf{Bespoke Fabric Flow}: This phase tailors a bespoke eFPGA fabric for each candidate module cluster. It identifies the fabric configuration that can host the candidate cluster yielding the lowest area overhead.  

    \item \textbf{eFPGA Selection Flow}: This phase selects the eFPGAs (a number decided by the user) that best meet our objective (maximize redacted logic within our area constraint) with no overlapping modules. 
\end{enumerate}
Ultimately, we integrate the selected eFPGAs into the design, replacing the redacted instances with the corresponding eFPGA ones and rerouting the signals as needed. The final design and the eFPGA netlists can be given to physical design tools.

\subsection{Secure Fabric Identification\label{sec:sec_identification}}

This pre-step needs to be performed offline once to identify a set of secure fabric parameters on which to perform design space exploration. Alternatively, one can look at the state of the art to identify these parameters. A paradigmatic study on the security of fabric configurations was conducted in~\cite{bhandari2021fabrics}.

In this step, the designers need to identify the current state-of-the-art attacks that could be used to evaluate the security of the fabrics. Each fabric configuration needs to be evaluated against the identified attacks.  This step is time-consuming but only needs to be performed once and eventually updated if new attacks are proposed. In general, the eFPGA fabrics that are compromised in this analysis will be excluded from the subsequent steps and the others can be explored and used in the rest of the flow. 

The results of this step might show that some fabric configurations are secure only above certain eFPGA sizes. If that is the case, these configurations should be considered in the step and discarded at the end only if the yielded size is not secure. 
An ensemble of multiple and diverse attacks can be used in this phase to identify the most resilient ones. However, in some cases, the designers can decide to keep some unsafe solutions with additional security measures. For example, suppose a final solution result is susceptible to SAT attacks but has minimal overhead. The designers might keep it and integrate a secure scan chain protection technique like DisORC~\cite{disorc} to protect against SAT attacks. Obviously, the overall overhead should be considered when evaluating the solution.

\Com{
Our redaction flow is shown in Figure \ref{fig:flow}. It starts from the RTL description of the design to be redacted in Verilog\footnote{Limitations are only due to the HDL parser we use. Supporting another HDL language (e.g., VHDL) only requires the proper parser.} and a set of parameters for the flow (in a custom YAML configuration file). Such parameters include eFPGA fabric configurations (e.g., as specified in the OpenFPGA configuration file), the maximum number of eFPGAs to be instantiated, and the maximum number of I/O for each of them. The number of I/O pins is also a rough indication of the type of eFPGA the designer aims to use. For example, a 4$\times$4 fabric configuration has no more than 64 I/O pins~\cite{our_iccad_21,mohan_hardware_2021}. 
Currently, we support (one or more) eFPGAs with the identical fabric architecture and maximum number of used I/O pins. While we contend that this setup will create a more regular physical design, adding support in the future for eFPGAs with different configurations is possible. 

ALICE focuses on how to partition an RTL design and generate the corresponding eFPGA-enhanced IC with three main phases: \textbf{module filtering}, \textbf{cluster identification}, and \textbf{eFPGA selection}. During module filtering, we analyze the design to identify \textit{candidate redaction modules}, while discarding the ones that do not satisfy specific constraints. In the second phase, the candidate modules are clustered into \textit{candidate module clusters}. Again, solutions that do not satisfy specific constraints are discarded. This phase results in a set of candidate clusters that are then characterized by running the flow to create the corresponding eFPGA fabrics. We finally apply an algorithm to select the eFPGAs that maximize our objectives (i.e., minimum hardware overhead and maximum security) with no overlapping sets of redacted modules. The resulting redacted RTL description is reproduced along with the fabrics of the selected eFPGAs. The final output is the description of the final system, which is ready for ASIC design.
ALICE is a modular flow that can be extended with additional criteria for selection. It can also interface with other eFPGA tools for characterization and include further metrics for security assessment if needed.
}



\subsection{Module Filtering}\label{sec:selection}

This stage is dedicated to analyzing the initial design to determine which RTL modules are candidates for redaction. The pseudocode for this procedure is outlined in \autoref{alg:filtering}. It starts with the initial RTL design $D$, a set of eFPGA parameters $P$ (i.e., the upper limit on the number of I/O pins), and a list of target outputs $O$. The algorithm then employs two distinct sets of criteria—\textit{functional} and \textit{structural}—to determine the ultimate list $R$ of modules to be redacted.
\textit{Functional} criteria aim to identify those modules that are more important for FPGA redaction from the functionality viewpoint, like the modules that directly affect the outputs in $O$. On the other hand, \textit{structural} criteria aim to spot modules suitable for eFPGA implementation while eliminating those that would render the design unworkable, like modules that exceed the I/O limit in $P$. This balanced approach ensures that the selected modules are crucial for the desired functionality and compatible with the constraints and capabilities of eFPGA technology.



The algorithm begins by looking at the functional criteria. We enumerate the modules $M$ in the initial design $D$ (line~\ref{l:extractmodules}), setting a starting score of zero for each (lines~\ref{l:initialize_score_b}-\ref{l:initialize_score_e}). We then construct the dataflow graph that captures the overall architecture of the RTL design. Subsequently, for every primary output in the target output list $O$, we update the scores of modules that exert a direct influence on that specific output (lines~\ref{l:score_b}-\ref{l:score_e}). Finally, modules with the highest scores are included in the list $F$, which comprises modules that are functionally significant for redaction (line~\ref{l:ranking}).


\begin{figure}[htb]
    \centering
    \parbox{0.75\textwidth}{%
\begin{algorithm} [H]
    
	\DontPrintSemicolon 
	\KwIn{Input RTL design $D$, eFPGA parameters $P$, list of selected outputs $O$}
	\KwOut{Set of candidate redaction modules $R$}
    $M \gets \Call{ExtractInitialModules}{D}$\tcp*{Analyze input RTL design.}\label{l:extractmodules}
	$S \gets \emptyset$\;
    \ForEach {$m \in M $}{\label{l:initialize_score_b}
       $S[m] \gets 0$\;
    }\label{l:initialize_score_e}
    \ForEach {$o \in O $}{\label{l:score_b}
       $T \gets \Call{IdentifyModules}{M, o}$\tcp*{Compute modules $T$ affecting $o$}    

	   $\Call{UpdateScore}{T, S}$\tcp*{Increment scores of modules $T$}    
    }\label{l:score_e}
    $F \gets \Call{RankAndSelect}{M, S}$\tcp*{Select most relevant modules}\label{l:ranking}
    $R \gets \emptyset$\;
	\ForEach {$f \in F $}{\label{l:structural_b}
    	\uIf{$\Call{CheckParameters}{f, P}$}{\label{l:check_param}
    		$R \gets R \cup \{ f \}$\;\label{l:add_list}
    	}
    }\label{l:structural_e}
    \Return{$R$}	\;
    
	\caption{ARIANNA module filtering}
	\label{alg:filtering}

\end{algorithm}}
    \end{figure}

In the subsequent phase, structural criteria are employed on each of the modules identified as functionally relevant for redaction (lines~\ref{l:structural_b}-\ref{l:structural_e}). Each module's compatibility with the specified eFPGA parameters is assessed (line~\ref{l:check_param}). For instance, we calculate the module's number of I/O pins to evaluate its fit within the prospective eFPGA fabric. Modules that meet these structural conditions are then added to the list $R$ (line~\ref{l:add_list}).

The list $R$ encompasses modules that not only influence a significant number of selected outputs but also meet the criteria for feasible eFPGA implementation. They can either be clustered together or stand-alone within an eFPGA fabric based on their dimensions. This step in the process is designed for adaptability, allowing for the inclusion of additional module-level filtering criteria as needed.








\subsection{Cluster Identification}\label{sec:identification}
\Com{
\textbf{Definition}: a module graph is a directed graph whose nodes are modules and whose edges represent instancing relationships between modules. If there is an edge from node A to node B, then module A creates an instance of module B. Consider a directed graph DG; DG is a module graph if each node represents a module and for each node N, for all modules M instantiated by module N, M is in DG. 

\textbf{Definition}: a cluster of modules is a disconnected graph whose subgraphs are all module graphs. 

\textbf{Clustering problem statement}: Given a set of candidate modules, a constraint on the eFPGA IO count, and a number of eFPGAs to be used, find all the sets of distinct clusters that respect the constraint, the number of clusters in the set must be less or equal to the number of eFPGA and each cluster must respect the constraint on the eFPGA. Two clusters A and B are distinct \cp{?? please clarify} \revision{if there does not exist an isomorphism between A and B} \lc{not sure if this is sufficient, but this would be concise}.
 
}
This stage focuses on identifying feasible combinations, called ``clusters,'' that can be redacted onto an eFPGA. A cluster can include a single module (\textit{single-module redaction}) or independent modules (\textit{multi-module redaction}). A cluster is deemed valid if its eFPGA implementation complies with the specified constraints established by the designer, ensuring the solution remains within the predefined parameters. 

\begin{figure}[t]
\centering
\includegraphics[width=.66\columnwidth]{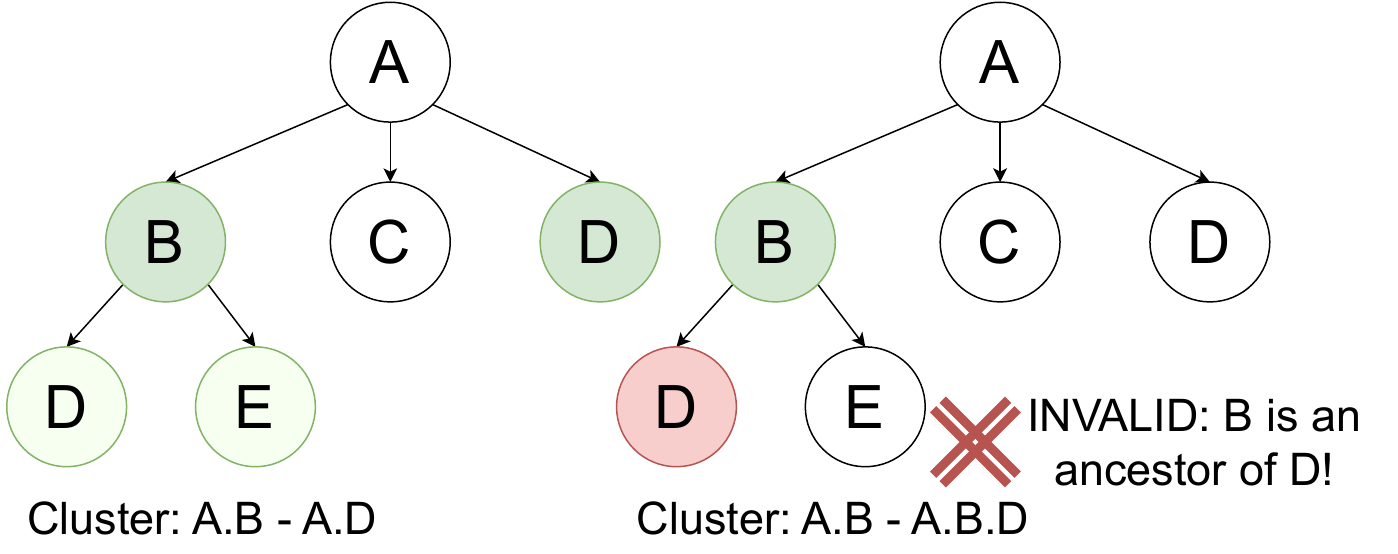}
\caption{Example of a valid and invalid cluster on a module hierarchy tree. Each node represents a module. An arrow from A to B means that module A instantiates module B. On the left, selecting B implicitly selects its sub-tree, including D and E. This is why, on the right, selecting both B and D explicitly yields an invalid cluster.}
\label{fig:ancestors}
\end{figure}

Two modules are allowed in the same cluster only if independent and not part of the same hierarchy. \autoref{fig:ancestors} shows a module hierarchy tree highlighting a valid and an invalid case. Although redacting a parent module without its children would be possible, that would require more I/O and data exchanges between the eFPGA and the non-redacted modules, increasing overheads.

\begin{figure}[htb]
    \centering
    \parbox{0.75\textwidth}{%
\begin{algorithm} [H]
\DontPrintSemicolon 
	\KwIn{Set of candidate redaction modules $R$, eFPGA parameters $P$}
	\KwOut{Set of candidate module clusters $C$}
	$C \gets \emptyset $\;
	\ForEach {$r \in R $}{\label{l:initialize_cluster_b}
       $C \gets C \cup \{ r \}$\;
    }\label{l:initialize_cluster_e}
    $Flag \gets False$\;
    \Do{Flag}{\label{l:fpa_b}
			$D \gets \emptyset$\; 
	       	\ForEach {$c1 \in C $}{\label{l:recombine_b}
         	\ForEach {$c2 \in C $}{
         	\If{$c1 \neq c2$}{
         		$N \gets c1 \cup c2$\; 
	       		\If {$N \not\subset D  \wedge N \not\subset C \wedge \Call{CheckParameters}{N, P}$}{\label{l:check_cluster_b}
	    			$D \gets D \cup N$\;\label{l:add_cluster}
    			}
    		}
    		}
    	}\label{l:recombine_e}
		$Flag \gets False$\;
    	\If{$D \neq \emptyset$}{\label{l:check_new_cluster_b}
    		$C \gets C \cup D$\;
    		$Flag \gets True$\;
    	}
    }\label{l:fpa_e}
    \Return{$C$}\;
	\caption{ARIANNA cluster identification}
	\label{alg:identification}
\end{algorithm}}
\end{figure}

\autoref{alg:identification} shows the pseudo-code for cluster identification performed by ARIANNA. Taking as input a set of candidate redaction modules $R$ and a set of eFPGA parameters $P$, it performs a \textit{fixed-point analysis} to identify the set $C$ of all candidate module clusters. Each of them is meant to fit into a single eFPGA; therefore, each cluster has to respect constraints dictated by the eFPGA parameters in $P$. 
We initialize the set $C$ with the trivial clusters, the ones composed of a single module identified in the previous phase (lines~\ref{l:initialize_cluster_b}-\ref{l:initialize_cluster_e}).
We then iteratively expand each cluster (lines~\ref{l:fpa_b}-\ref{l:fpa_e}). This part involves the recombination of each pair of admissible clusters to identify candidates for the current iteration (lines~\ref{l:recombine_b}-\ref{l:recombine_e}). Suppose the cluster was not already identified in the previous iterations and it respects the eFPGA constraints $P$ and hierarchy constraints (line~\ref{l:check_cluster_b}). In that case, it is considered a new valid cluster and added to the list of current clusters (line~\ref{l:add_cluster}).
Each cluster's evaluation follows the same structural criteria applied to individual modules. For instance, in the context of multi-module redaction, the total number of I/O pins is calculated by aggregating the I/O pins of the individual modules within the cluster. The cluster earns the status of being admissible if it adheres to the designer's predetermined constraints.
At the end of each iteration, the new clusters (line~\ref{l:check_new_cluster_b}) are added to the set $C$, and the procedure restarts. 
We terminate our algorithm when it is impossible to create new clusters by recombining the current ones.
At the end of the procedure, each element of $C$ is a candidate module cluster. 

If a cluster contains more than one module, a module wrapper must be created, as OpenFPGA expects a single module. The module wrapper instantiates all the modules in the cluster and exposes their I/Os through its ports.




\Com{Each element of the set $C$ is a list of candidate modules. We build the set $C$ iteratively. The procedure begins by inserting, in $C$ a list containing only $e \in M, \forall  e \in  M$ \$ . Then, for each element $c$ in $C$ we proceed by adding to $C$ all possible expansions of $c$ using the candidate modules in $M$. We add a new combination in $C$ only if it is not already present. A module $m$ can expand a combination $c$ if: (a) $\forall c' \in c, c'$ and $m$ are unrelated modules; (b) the combination obtained by adding m to c satisfies the constraints. If at least an expansion of $c$ has been found, we insert $c$ in an auxiliary set $D$ and remove it from $C$. We keep doing this expansion step until we do not expand any combination in an iteration. At the end of the procedure, we do a union between $C$ and $D$  to get all the candidate module combinations.}



\subsection{Bespoke Fabric Flow}\label{sec:heuristics}
This step aims to tailor a bespoke eFPGA fabric for each candidate module cluster identified in the previous phase.
As shown in \autoref{sec:fpga_arch}, OpenFPGA allows customization of CLBs, changing the number of LUTs (N) and the number of inputs for each LUT (K). We propose two design space exploration heuristics for finding $N$ and $K$ that will yield low area overhead. For both heuristics, we assume designers provide a range for $N$ and $K$ that they consider secure for their application, determined from the secure fabric identification step (\autoref{sec:sec_identification}). 

Specifically, we propose two heuristics whose object is to find the smallest $N$ and $K$ that will yield the minimum tile number. Both heuristics start by identifying a lower bound for the tile number with the given cluster candidate. This is done by running OpenFPGA using the largest $N$ and $K$ provided by the user. This configuration provides the biggest CLBs. Therefore, the number of CLBs will be minimal. If the number of tiles exceeds the allowed one, we discard the current cluster candidate. Otherwise, we start our search. 

The first heuristic minimizes $N$ such that the tile number does not change and then minimizes $K$. We refer to this heuristic as $NK$. The second heuristic minimizes $K$ such that the tile number does not change and then minimizes $N$, we refer to this heuristic as $KN$. A pseudo-code for the two heuristics is provided in \autoref{alg:nk_heu} and \autoref{alg:kn_heu}, respectively.
Both $NK$ and $KN$ heuristics begin by identifying the lower bound for the eFPGA size. To do this, we run OpenFPGA using the maximum $N$ and $K$ parameters provided by the user (line~\ref{l:min_size}). If the lower bound size exceeds the maximum allowed, we end the procedure, marking the cluster as unfeasible due to resource constraints (line~\ref{l:exit}. If the identified lower bound size is admissible, we proceed by iteratively decreasing $N$ (for $NK$ heuristic) or $K$ (for $KN$ heuristic) until the size obtained by the reduced parameter increases (lines~\ref{l:N1}-\ref{l:N2}). If this happens, we select the $N$ or $K$ value from the previous iteration; otherwise, we stop at the minimum value for the parameter. Once we end this first iteration, we proceed by doing the same to minimize the second parameter (i.e., $K$ for $NK$ and $N$ for $KN$ -- lines~\ref{l:K1}-\ref{l:K2}).
This procedure explores only the minimum size configurations, which are the fastest to compute by OpenFPGA. Thus, it allows us to reduce the number of OpenFPGA runs required and eliminate the most demanding ones.

The framework can be run with either heuristic to obtain a bespoke fabric for each candidate module cluster. Given a module and a fabric configuration, OpenFPGA returns the corresponding eFPGA if the design is feasible and an error otherwise (e.g., when the cluster modules cannot be implemented for any reason). As a consequence, this phase also filters out unfeasible module clusters. 

\begin{figure}[htb]
    \centering
    \parbox{0.75\textwidth}{%
\begin{algorithm} [H]
	\DontPrintSemicolon 
	\KwIn{Candidate module cluster C $D$, min N $n$, max N $N$, min K $k$, max K $K$, max size $S$}
	\KwOut{Optimal N $No$, optimal K $Ko$}
    $s \gets \Call{OpenFPGA}{C,N,K}$\tcp*{Get minimum size for current cluster}\label{l:min_size}
    \If{$s > S$}{\Return{$None$}\label{l:exit}} 
    \For{$i \in [1, N-n]$}{  \label{l:N1}
        $Si \gets \Call{OpenFPGA}{C,N-i,K}$\;
        \If{$Si > s$}{break \tcp*{Decrease K until size increases}}
        $No \gets N-i$ \label{l:N2}
    }
    \For{$i \in [1, K-k]$}{ \label{l:K1}
        $Si \gets \Call{OpenFPGA}{C,No,K-i}$\;
        \If{$Si > s$}{break \tcp*{Decrease K until size increases}}
        $Ko \gets K-i$\label{l:K2}
    }
    \Return{$No, Ko$}	\;
	\caption{NK Heuristic}
	\label{alg:nk_heu}
\end{algorithm}}
\end{figure}

\begin{figure}[htb]
    \centering
    \parbox{0.75\textwidth}{%
\begin{algorithm} [H]
	\DontPrintSemicolon 
	\KwIn{Candidate module cluster C $D$, min N $n$, max N $N$, min K $k$, max K $K$, max size $S$}
	\KwOut{Optimal N $No$, optimal K $Ko$}
    $s \gets \Call{OpenFPGA}{C,N,K}$\tcp*{Get minimum size for current cluster}
    \If{$s > S$}{\Return{$None$}}
    \For{$i \in [1, K-k]$}{ 
        $Si \gets \Call{OpenFPGA}{C,N,K-i}$\;
        \If{$Si > s$}{break \tcp*{Decrease K until size increases}}
        $Ko \gets K-i$
    }
    \For{$i \in [1, N-n]$}{ 
        $Si \gets \Call{OpenFPGA}{C,N-i,Ko}$\;
        \If{$Si > s$}{break \tcp*{Decrease K until size increases}}
        $No \gets N-i$
    }
    \Return{$No, Ko$}	\;
	\caption{KN Heuristic}
	\label{alg:kn_heu}
\end{algorithm}}
\end{figure}

\subsection{eFPGA Selection}\label{sec:efpga_selection}

At this stage, each candidate module cluster in $C$ has been associated with its bespoke eFPGA fabric. The resulting candidate implementations must be characterized, ranked, and selected to determine the final solution. In this phase, we evaluate the utilization, and all candidate clusters select the best and final ones--more than one if the user decides to use more than one FPGA). 

\begin{figure}[htb]
    \centering
    \parbox{0.75\textwidth}{%
\begin{algorithm} [H]
	\DontPrintSemicolon 
	\SetKwProg{Fn}{Function}{:}{}
  	\KwIn{Set of candidate module clusters $C$, eFPGA parameters $P$}
	\KwOut{Solution $s_t$}
	$F \gets \emptyset$\;
	\ForEach {$c \in C $}{\label{l:char_openfpga_b}
       $f \gets \Call{CreateEFPGA}{c, P}$\;\label{l:runopenfpga}
       \If{$\Call{IsValid}{f}$}{\label{l:checkvalid}
       	   $F \gets F \cup f $\;\label{l:keepefpga}
	   }
    }\label{l:char_openfpga_e}
    $T \gets \Call{ComputeScore}{F}$\;\label{l:scoreefpga}
    $W \gets \{ \}$\tcp*{Initialize with empty solution}\label{l:initialize_empty}
    $S \gets \emptyset$\;
    \ForEach{$w \in W$}{\label{l:bb_b}
    	\ForEach{$f \in F$}{\label{l:addnewefpga_b}
    		$c \gets f \cup w$\;\label{l:createsolution}
    		\If{$\Call{isValidSolution}{c}$}{\label{l:checkvalidsol}
    			\If{$\Call{isFinal}{c}$}{\label{l:ifleaf}
    				$S \gets S \cup c$\;\label{l:addsolution}
    			}
    			\Else{
    				$W \gets W \cup c$\;\label{l:addworking}
    			}
    		}
    	}\label{l:addnewefpga_e}
    }\label{l:bb_e}
    $S \gets S \cup W \setminus \{ \}$\;
    $s_t \gets \Call{RankAndSelect}{S, T}$\;\label{l:rankselect}
    \Return{$s_t$}\;
	\caption{ARIANNA eFPGA selection}
	\label{alg:selection}
\end{algorithm}}
\end{figure}

In \autoref{alg:selection}, we provide the pseudocode for this phase, detailing the step-by-step procedure. Initially, we parse the logs generated by OpenFPGA for the customized fabrics of each candidate module cluster, extracting essential data on CLB and I/O utilization (line~\ref{l:scoreefpga}). Subsequently, we compute a score for each fabric implementation, considering information regarding both I/O and CLB utilization as follows:
\begin{equation}\label{eq:score}
	T_f = \frac{\mbox{MaxIOUtil} - \mbox{IOUtil}_f}{\mbox{MaxIOUtil}}+\frac{\mbox{MaxCLBUtil} - \mbox{CLBUtil}_f}{\mbox{MaxCLBUtil}}
\end{equation}
where: 
\begin{itemize}
    \item  $\mbox{IOUtil}_f$ and $\mbox{CLBUtil}_f$ represent the I/O and CLB utilization, respectively.
    \item $\mbox{MaxIOUtil}$ and $\mbox{MaxCLBUtil}$ represent the corresponding maximum I/O and CLB values for all analyzed eFPGAs, respectively.
  
\end{itemize}
This scoring approach balances the use of I/O and CLB resources and incorporates considerations related to security resilience. Fabrics with lower I/O utilization are generally more susceptible to certain attacks, as they can potentially reveal stuck-at-0 outputs more easily. Similarly, fabrics with lower CLB utilization provide less logic to be successfully recovered, further contributing to security resilience. These aspects are integral to our comprehensive evaluation framework, as described in our previous works~\cite{our_iccad_21,bhandari2021fabrics}.
%

In our approach, we employ a \textit{branch\&bound algorithm} to systematically enumerate all possible combinations of eFPGAs that can be redacted together (lines~\ref{l:bb_b}-\ref{l:bb_e}), resulting in a comprehensive set of solutions. The algorithm begins with an empty working solution (line~\ref{l:initialize_empty}) and, at each iteration, strives to incorporate a new eFPGA implementation into each existing working solution (lines~\ref{l:addnewefpga_b}-\ref{l:addnewefpga_e}). Here, a solution denotes a collection of eFPGAs with non-overlapping module instances. If a solution reaches a terminal state (i.e., it either reaches the maximum allowable eFPGAs or redacts all eligible modules), it is appended to the final set of solutions (line~\ref{l:addsolution}). Otherwise, it remains in the working list for potential expansion (line~\ref{l:addworking}). Upon the conclusion of this phase, the set $S$ encompasses the complete assortment of viable solutions.
Subsequently, we proceed to calculate a score for each solution. The score of a solution is derived as the summation of the scores of its constituent eFPGA implementations, with each score determined using \autoref{eq:score}. The set $S$ is then ranked based on these scores, with the highest-scoring solution being designated as the best and ultimate solution (line~\ref{l:rankselect}). This rigorous evaluation and ranking process enable us to identify the most optimal redaction solution from the pool of candidates.



In the final solution, we assemble a set of eFPGA implementations, each comprising a roster of module instances to be redacted. At this juncture, our next task is to regenerate the top module for ASIC implementation (referred to as the ``Top ASIC module'' in \autoref{fig:flow}). This involves substituting the redacted instances with their corresponding eFPGA instances. In scenarios involving multi-module redaction, where multiple modules may be distributed throughout the design, we conduct a ``dominator tree'' analysis on the module hierarchy. This analysis helps identify the optimal insertion points for eFPGA instances, intending to minimize wire lengths.

During this process, we re-route signals originating from the original instances to their respective eFPGA instances. Additionally, control signals are propagated to the top module as required. We also remap the module signals to correspond with the eFPGA GPIO (General Purpose Input/Output) signals to ensure correct connectivity.

Once these modifications are complete, the updated design and fabric netlists are ready for handoff to physical design tools, facilitating the translation of the design into a finalized, manufacturable ASIC implementation. This comprehensive approach ensures that the redaction process seamlessly integrates eFPGA solutions into the ASIC design, optimizing functionality and security.





\section{Experimental Evaluation}\label{sec:results}

We implemented a prototype of ARIANNA in Python, using the PyVerilog framework~\cite{Takamaeda:2015:ARC:Pyverilog}. PyVerilog can parse the Verilog designs, analyze and manipulate the resulting Abstract Syntax Tree (AST), and regenerate the output files, including those fed into the OpenFPGA toolchain for eFPGA creation.

To identify the set of secure fabrics, we run a security evaluation considering IcySAT~\cite{shamsi_icysat_2019} as the reference attack. We selected IcySAT as it is the most powerful attack with an open-source implementation that we were able to find.

Then, we conduct an exhaustive analysis of the design space of fabric parameters for each benchmark. This preliminary analysis helped us formulate the proposed heuristics and allowed us to show the complexity of the problem. We show the results obtained with the proposed heuristics for the bespoke fabrics, comparing them with the exhaustive approach. Eventually, we show results from the complete flow using Cadence Genus 18.14 for logic synthesis and Cadence Innovus 18.10 for physical design, targeting the NanGate 45nm Open Cell Library.

\autoref{tab:bench} shows the benchmarks we used to validate ARIANNA.  
The table reports the number of modules and instances that can be redacted. We report the range of the I/O pin count for such modules. We identified the main data output(s) of each design as the outputs of interest for the module filtering phase.
These benchmarks are commonly used to evaluate RTL locking~\cite{9427060}. In the context of eFPGA redaction, we can consider these designs as IPs part of a bigger SoC. The designer wants to protect this IP, but redacting the whole IP is not feasible. Using ARIANNA, the designer can find redaction solutions that satisfy the constraints.

\begin{table}[t]
\caption{Characteristics of the selected benchmarks}\label{tab:bench}
\centering
\vspace{-8pt}\begin{tabular}{@{}llcccl@{}} 
\toprule
\textbf{Suite}                   & \textbf{Design}   & \textbf{Modules}  & \textbf{Instances} & \textbf{I/O pins}  \\ 
                   &    & \textbf{(\#)}   & \textbf{(\#)} & \textbf{[min, max]}  \\ 
\midrule
\multirow{4}{*}{CEP~}   & DES3     & 11 & 11 & [12, 301]       \\
                        & FIR      & 5  & 5 & [64, 384]         \\
                        & SHA256   & 3  & 3 & [38, 774]        \\ 
\midrule
\multirow{2}{*}{IWLS05} & SASC     & 2  &  3 & [23, 28]        \\
                        & USB\_PHY & 3  & 3 & [17, 33]      \\ 
\midrule
OpenROAD & GCD    & 10  & 11 & [6, 68]         \\
\midrule 
Opencores & Nautilus & 8 & 8 & [19, 93]  \\
\bottomrule
\end{tabular}
\end{table}

\subsection{Secure Fabric Identification\label{sec:exp_sec}}
For the secure fabric identification pre-step, we considered IcySAT~\cite{shamsi_icysat_2019} as the reference attack. IcySAT is the most powerful attack with an open-source implementation that we could find. 
We ran IcySAT on each fabric configuration with a timeout of 48 hours.
Designers might choose a different attack or multiple attacks and limit or increase the eFPGA size they are interested in and the attack(s) timeout.

To run the attack, we must convert the gate-level netlist into a format that can be understood by an attack tool, treating the configuration bitstream as a collection of ``key inputs.'' In an eFPGA, the bitstream is stored in configuration flip-flops. These configuration flip-flops are linked together in a scan chain that is controlled by a programming clock (prog\_clk). To locate the configuration scan chain, we perform a depth-first search of the netlist, beginning from the scan\_in\_head port and continuing until we arrive at the scan\_in\_tail.
Every flip-flop (FF) in the traversal path influenced by the programming clock (prog\_clk) retains the configuration bitstream. The sequence in which the configuration FFs are identified aligns with the order of the bitstream. The recognized configuration FFs serve as primary key inputs to transform the eFPGA netlist into a version compatible with IcySAT. To create an Oracle, we utilize the same locked netlist but assign the key bits the configuration values derived from the bitstream produced in the OpenFPGA process.

\begin{figure*}[htbp]
    \centering
    \begin{tabular}{ccc}
        \hspace{-17pt}\includegraphics[width=0.35\textwidth]{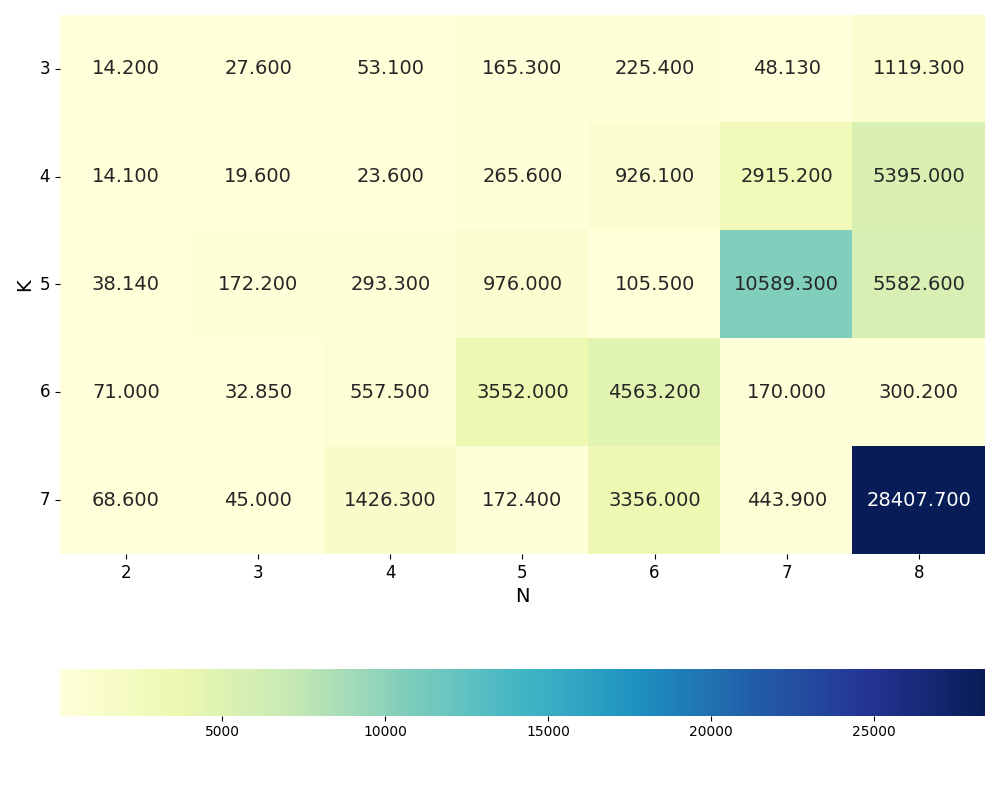} &
        \hspace{-14pt}\includegraphics[width=0.35\textwidth]{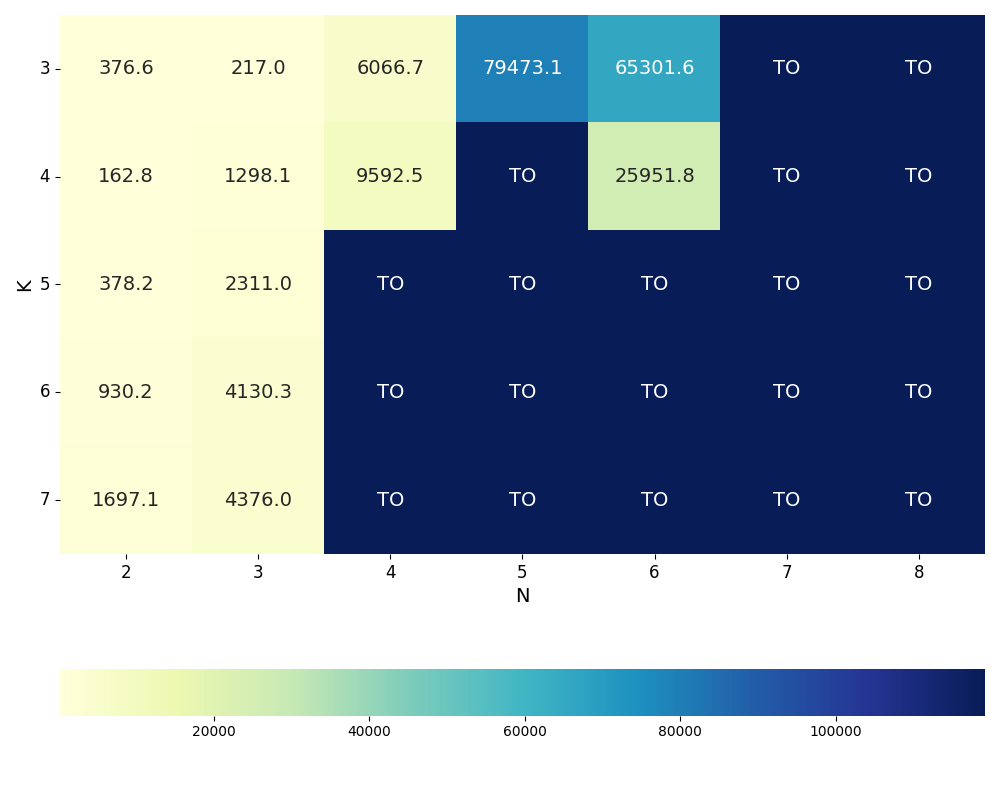} &
        \hspace{-14pt}\includegraphics[width=0.35\textwidth]{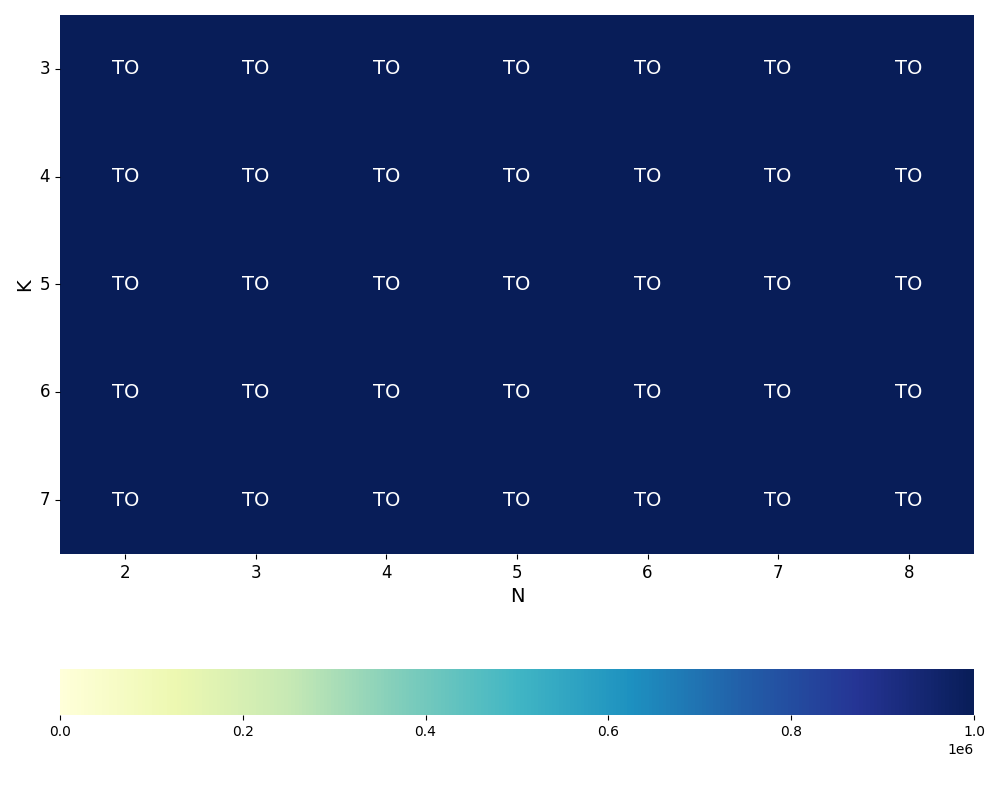} \\
a) 4x4&b) 5x5&c) 6x6 \\
    \end{tabular}
    \vspace{-10pt}
    \caption{SAT attack results [s]. Timeout is 48 hours.}
    \label{fig:sat}
\end{figure*}

\autoref{fig:sat} shows the results for each fabric configuration for a 4x4, 5x5, and 6x6 eFGPA size.
The results show that for 4x4 eFPGA size, all attacks run to completion before timeout. For a 5x5 eFGPA size, only the smaller tiles get to completion before the timeout. For 6x6 eFPGA size, all fabrics stop at timeout. From this, all fabrics should be considered as all would be safe from IcySAT if the final eFPGA size is 6x6 or bigger.

\subsection{Heuristics Evaluation} \label{sec:exh}
\begin{figure*}[htbp]
    \centering
    \begin{tabular}{ccc}
        \hspace{-17pt}\includegraphics[width=0.35\textwidth]{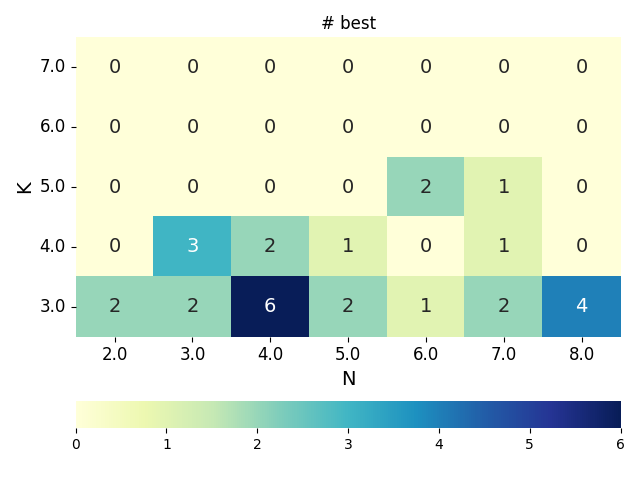} &
        \hspace{-14pt}\includegraphics[width=0.35\textwidth]{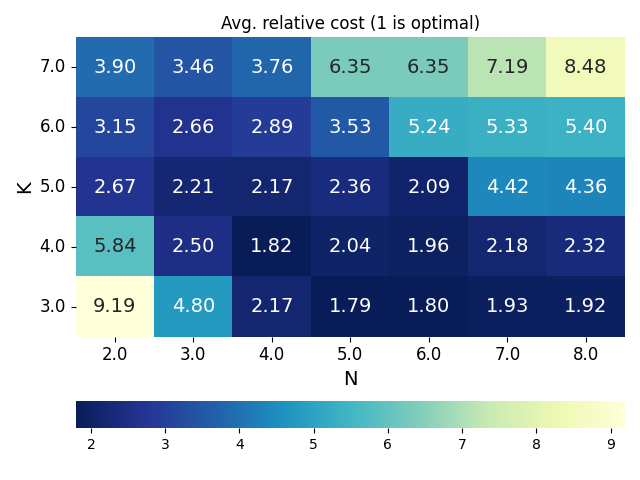} &
        \hspace{-14pt}\includegraphics[width=0.35\textwidth]{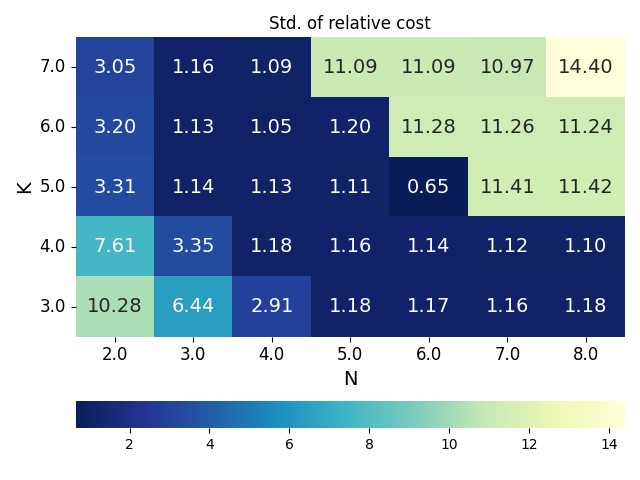} \\
       \hspace{-17pt}\includegraphics[width=0.35\textwidth]{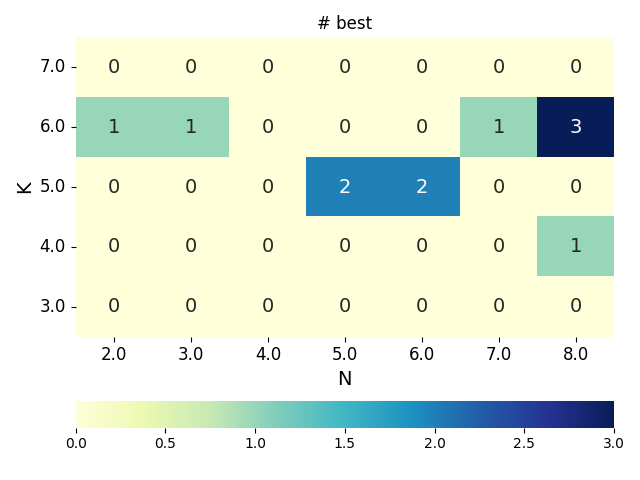} &
       \hspace{-14pt}\includegraphics[width=0.35\textwidth]{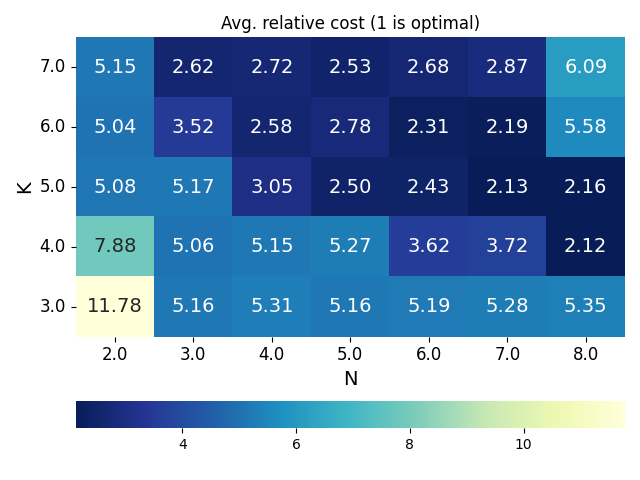} &
       \hspace{-14pt}\includegraphics[width=0.35\textwidth]{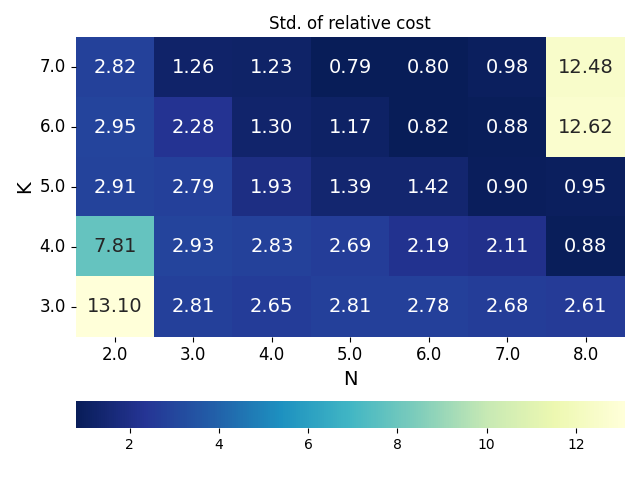} \\
        \hspace{-17pt}\includegraphics[width=0.35\textwidth]{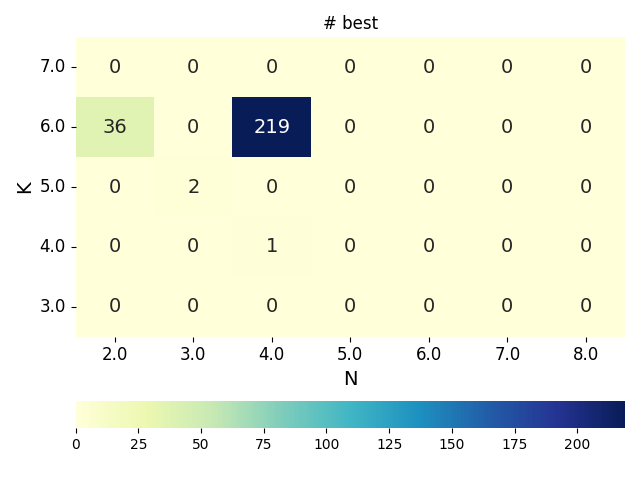} &
        \hspace{-14pt}\includegraphics[width=0.35\textwidth]{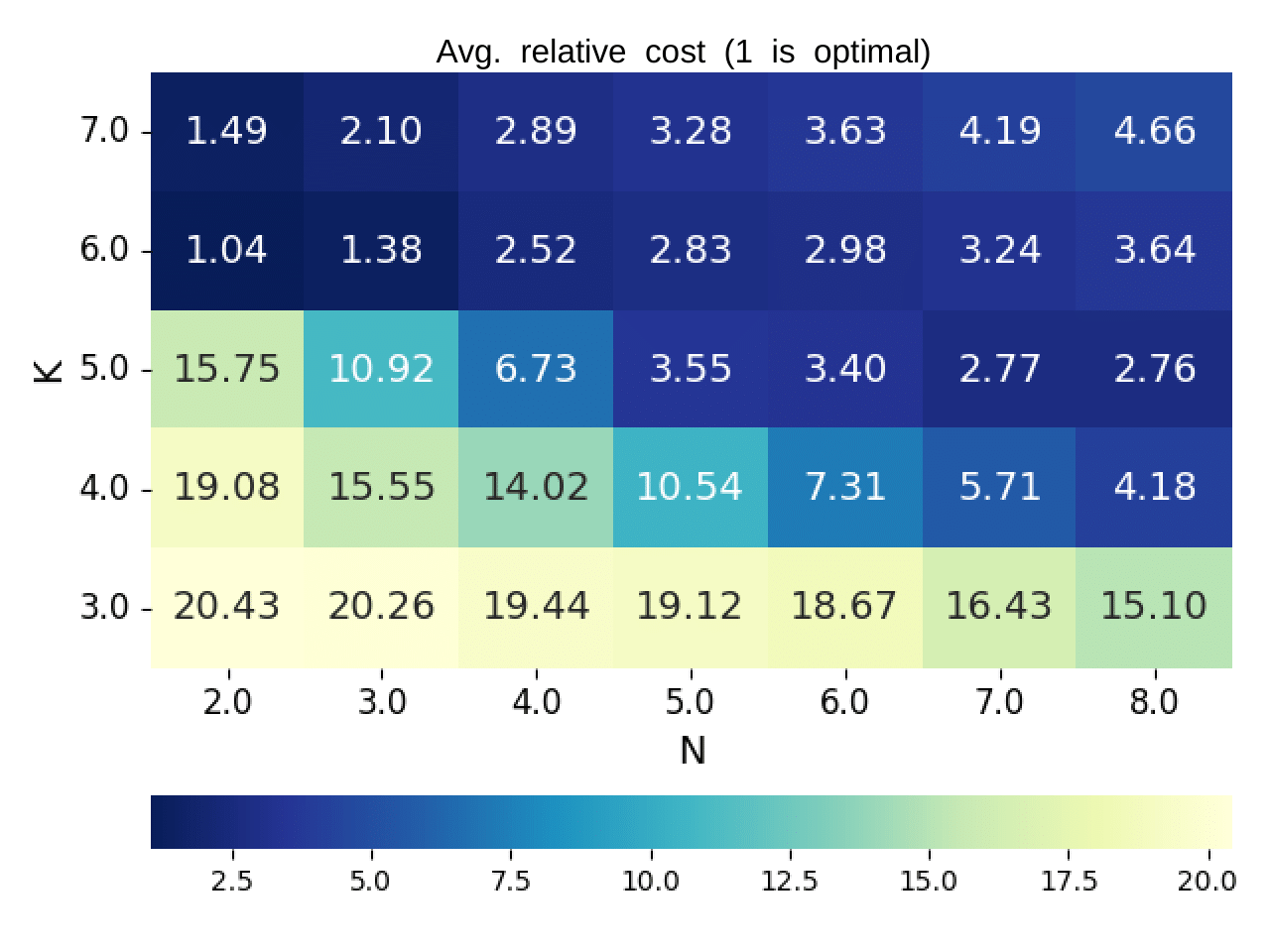} &
        \hspace{-14pt}\includegraphics[width=0.35\textwidth]{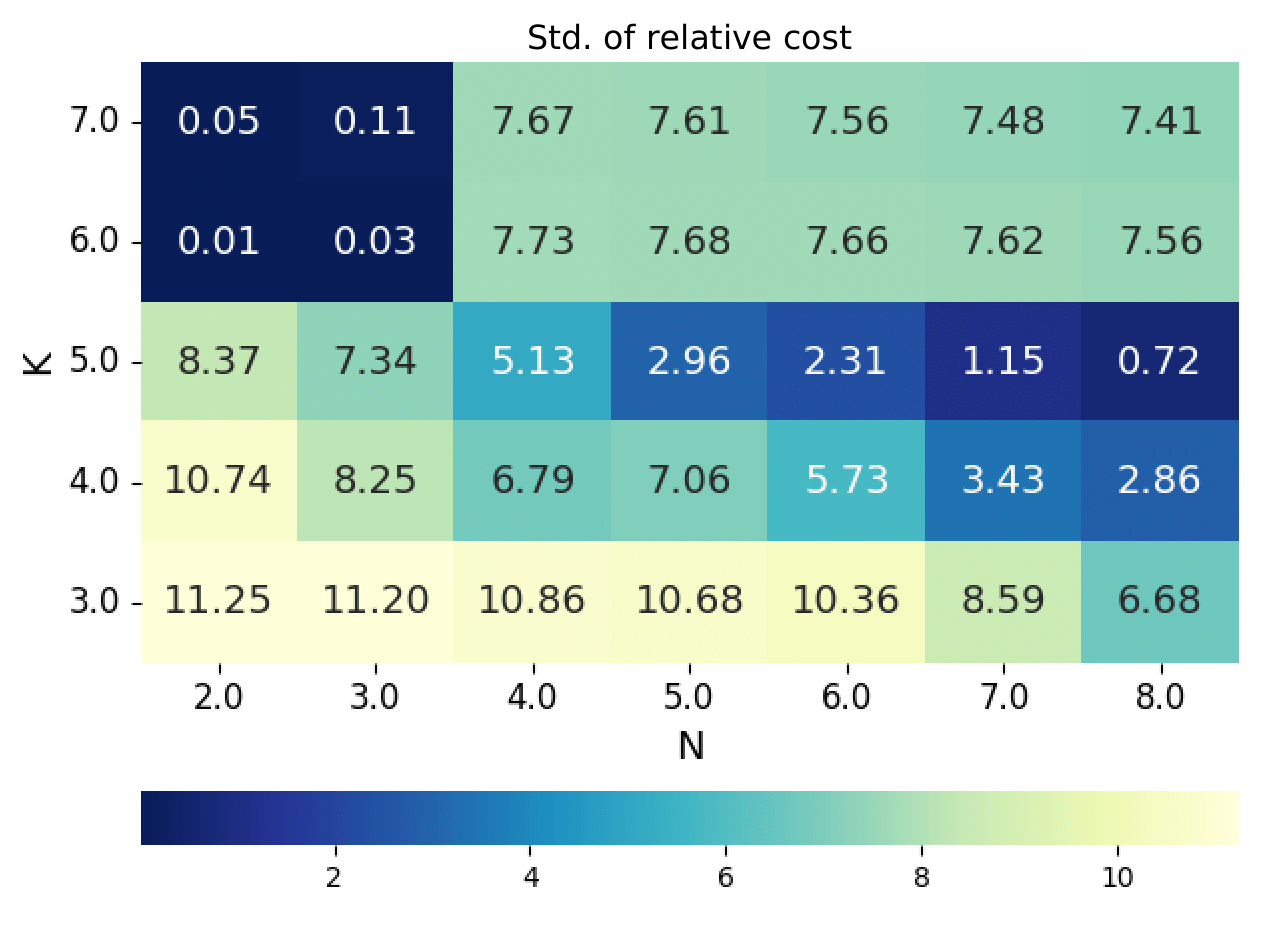} \\
        
    \end{tabular}
    \caption{Number of best results, average and standard deviation of distance from the best result for GCD (top row), Nautilus (middle row), and DES3 (bottom row) for each fabric configuration (given by the N-K pair). Darker is better. There is not a predominant configuration. The most frequent best solution for DES3 has a high average relative cost, highlighted by its std. Motivating the need for heuristics to search the space.}
    \label{fig:exhaustive_res}
\end{figure*}

In our preliminary analysis, we ran an exhaustive search for each candidate module cluster and saved all the results with the different fabric configurations. Overall candidate module clusters, overall benchmarks (291 in total), and the number of times an FPGA with a nonminimum number of tiles has the minimum area equal to 22. 
\autoref{fig:exhaustive_res} shows how many times each fabric configuration (identified by the N-K pair) yields the best result, their average relative cost (with respect to the best result), and the standard deviation of the relative cost for GCD, Nautilus, and DES3. 
We show these three benchmarks as they are more meaningful because they have a larger number of candidate module clusters compared to the rest of the benchmarks. From the heatmaps, we can see two different cases. For GCD (top row), no predominant fabric configuration is always good, especially as highlighted by the average relative distance from the best. For DES3 (bottom row) we can see that the most predominant fabric configuration (K6 and N4), is not very good on average (250\% average relative cost), this is reflected by the high standard deviation, indicating that this fabric is either very good or very bad. Conversely, the K6 N2 configuration has a very low average relative cost (104\%) and very low standard deviation, indicating that this configuration is always pretty good, although not very frequently the best. Though, we must notice that is always pretty good for the candidate module clusters in this design, for GCD that configuration has an average relative cost of 252\%. This motivates the tailoring of a bespoke fabric for each candidate module cluster. As the number of best solutions that do not have minimum tile numbers is marginal, we search for the solutions with minimum tile numbers. See \autoref{sec:heuristics} for a detailed explanation of the proposed heuristics.

We ran the redaction flow to evaluate the proposed heuristics and focused on the bespoke fabric results for each candidate cluster. We compared the results of the two heuristics against the exhaustive analysis results shown in \autoref{sec:exh}. \autoref{tab:heu} summarizes the results, reporting the average and standard deviation of the relative cost and the computation time. The two heuristics perform very well in all benchmarks, with GCD being an outlier. The two heuristics present very close computation times, often over an order of magnitude lower than the exhaustive approach. In fact, as the heuristics start exploring the bigger tiles and stop at the first iteration in which the grid size increases, they do not only cut the number of explored solutions, but they also explore the least expensive ones in terms of computations. This is because the computation time of OpenFPGA increases as the number of tiles increases. 

\begin{table}[t]
\centering
\caption{Heuristic vs Exhaustive approach. The cost is relative to the best solution found with the exhaustive search (the best solution has cost 1).
}
\label{tab:heu}
\begin{tabular}{@{}lrrrrrrrrr@{}}
\toprule
\multicolumn{1}{l}{\multirow{2}{*}{Benchmark}} & \multicolumn{2}{c}{\# best} & \multicolumn{2}{c}{Avg. Rel. Cost} & \multicolumn{2}{c}{Std. Rel. Cost} & \multicolumn{3}{c}{Time [s]} \\
\multicolumn{1}{c}{} & \multicolumn{1}{c}{NK} & \multicolumn{1}{c}{KN} & \multicolumn{1}{c}{NK} & \multicolumn{1}{c}{KN} & \multicolumn{1}{c}{NK} & \multicolumn{1}{c}{KN} & \multicolumn{1}{c}{NK} & \multicolumn{1}{c}{KN} & \multicolumn{1}{c}{Exh.} \\ \midrule
DES3 & 247 & 222 & 1.0 & 1.14 & 0.0 & 0.45 & 5962 & 5453 & 31170 \\
FIR & 0 & 1 & 1.56 & 1.03 & 0.46 & 0.03 & 58 & 53 & 968 \\
SHA & 1 & 1 & 1.0 & 1.0 & 0.0 & 0.0 & 13 & 10 & 162 \\
SASC & 0 & 1 & 1.53 & 1.0 & 0.0 & 0.0 & 15 & 14 & 86 \\
USB      & 0 & 0 & 1.04 & 1.33  & 0.0 & 0.0 & 23 & 21 & 373 \\ 
GCD & 2 & 11 & 4.78 & 4.36 & 13.61 & 13.70 & 289 & 287 & 1986 \\
Nautilus & 6 & 7 & 1.21 & 1.07 & 0.27 & 0.14 & 277 & 274 & 5653 \\
\bottomrule
\end{tabular}%

\end{table}

\subsection{Framework Evaluation}\label{sec:eval}

To evaluate the complete flow, we set ARIANNA to run in two configurations for each heuristic and compared our results with the previous work~\cite{alice} where N and K were set to 4. We run two configurations to see the effects of redacting with a single but bigger eFPGA versus redacting with two smaller eFPGAs.  In \texttt{cfg1}, we set the maximum I/O pin count of the modules that can redacted to 64, and the limit is two eFPGAs. In \texttt{cfg2}, the maximum I/O pin count is 96, and the limit is one eFPGA. 
The results of the Clustering Flow (\autoref{sec:identification}) are going to give the same inputs to the two heuristics, which will potentially find different solutions for the bespoke fabrics (\autoref{sec:heuristics}) of the candidate clusters, leading to different scores in the final selection phase (\autoref{sec:efpga_selection}).
We validated the designs with Cadence Genus 18.14 for logic synthesis, targeting the NanGate 45nm Open Cell Library.

\begin{table*}[]
\caption{Flow results after running ARIANNA with two different configurations, with the previous work~\cite{alice} (prev.) and the proposed heuristics.}\label{tab:results}
\resizebox{\textwidth}{!}{%
\begin{tabular}{@{}llrrrrrrrrrrrrrr@{}}
\toprule
Config. &
  \multicolumn{1}{c}{Design} &
  \multicolumn{1}{c}{\# Instances} &
  \multicolumn{2}{c}{Module Filtering} &
  \multicolumn{2}{c}{Cluster Ident.} &
  \multicolumn{9}{c}{Fabric Exploration \& eFPGA Selection} \\ \cmidrule(l){4-5}\cmidrule(l){6-7}\cmidrule(l){8-16} 
 &
  \multicolumn{1}{c}{} &
  \multicolumn{1}{c}{} &
  \multicolumn{1}{c}{Time [s]} &
  \multicolumn{1}{c}{$\|R\|$} &
  \multicolumn{1}{c}{Time [s]} &
  \multicolumn{1}{c}{$\|C\|$} &
  \multicolumn{3}{c}{Time [s]} &
  \multicolumn{3}{c}{\# OpenFPGA runs} &
  \multicolumn{3}{c}{$\|S\|$} \\
  \cmidrule(l){8-10}\cmidrule(l){11-13}\cmidrule(l){14-16} 
                                               &          &  &  &  &  &  & ALICE & NK & KN & ALICE & NK & KN & ALICE & NK & KN \\ \midrule
\multirow{7}{*}{\rotatebox[origin=c]{90}{\makecell{cfg1:\\64 I/O bits\\\& 2 eFPGAs}}} & DES3 & 11 & 338.7 & 8 & 1.2 & 218 &  748.4  & 4383.1 & 4210.3 & 218 & 1716 & 1632 & 3151 & 2219 & 2219\\
                                               & FIR      & 5 & 0.4 & 1 & 0.0 & 1 & 2.1 &  19.8  &  16.1  &   1    &  8  &  7  &  1     &  1  &  1  \\
                                               & SHA256   & 3 & 15.3 & 1 & 0.0 & 1 &    5.3   &  11.3  &  11.0  &   1    &  4  &  4 &   1    &   1 &  1  \\
                                               & SASC     & 3 & 0.3 & 1 & 0.0 & 1 & 2.8      &   15.1 &  13.9  &   1    &  6  & 6   &   1    &  1  &  1  \\
                                               & USB\_PHY & 3 & 1.2 & 3 & 0.0 & 3 &  9.7     & 21.3   & 21.1   &    3   &  7  & 7   &    1   &  1  & 1   \\
                                               & GCD      & 11 & 0.6 & 8 & 0.0 & 17  &     32.5  &  193.9  &  183.2  &     17  &  96  & 98   &    78   &   69 &  69  \\
                                              & NAUTILUS & 8 & 104.1 & 3 & 0.0 & 6 &    36.0   &   95.4 &   99.7 &   6    &   24 &    &   6    &  6  &  6  \\ 
                                              \midrule
\multirow{7}{*}{\rotatebox[origin=c]{90}{\makecell{cfg2:\\96 I/O bits\\\& 1 eFPGA}}} & DES3 & 11 & 336.7 & 8 & 1.3 & 255 & 861.3 & 5356.0 & 5454.1 & 255 & 2012 & 1928 & 232 & 247 & 247 \\
                                               & FIR      & 5 & 0.2 & 3 & 0.0 & 3 &    29.3   &  57.1  &  59.5  &   3    &  16  &  15  &   3    &  2  &  2  \\
                                               & SHA256   & 3 & 15.1 & 1 & 0.0 & 1 &     5.3  &  10.5  &  10.4  &   1    &   4 & 4   &  1    &  1  &  1  \\
                                               & SASC     & 3 & 0.3 & 1 & 0.0 & 1 &  2.7     &   13.9 &  14.0  &    1   &  6  &   6 &   1    & 1   &  1  \\
                                               & USB\_PHY & 3 & 1.2 & 3 & 0.0 & 3  &   9.5    &  20.8  &   23.7 &     3  &  7  &  7  &    1   &  1  &  1  \\
                                               & GCD      & 11 & 0.6 & 9 & 0.1 & 43 &     94.3  &  527.2  &  287.2  &     43  &   238 &  141  &   33    &   32 &  19  \\
                                               & NAUTILUS & 8 & 134.6 & 5 & 0.0 & 9 &    73.4   & 181.8   &  99.7  &    9   &  35  & 23   &    6   &  5  &   5 \\
                                               \bottomrule
\end{tabular}
}
\end{table*}

\begin{table*}[]
\caption{Redaction results after running ARIANNA with two different configurations with the previous work~\cite{alice} (prev.) and the proposed heuristics. Redacted module lists are separated by ``;'' to indicate different eFPGAs. In all cases, the I/O utilization is improved using the novel heusristics for bespoke fabrics.
}\label{tab:results2}
\resizebox{\textwidth}{!}{%
\begin{tabular}{@{}lllllccccccrrrrrr@{}}
\toprule
Config. &
  \multicolumn{1}{c}{Design} &
  \multicolumn{3}{c}{\begin{tabular}[c]{@{}c@{}}Redacted\\ Modules\end{tabular}} &
  \multicolumn{3}{c}{\begin{tabular}[c]{@{}c@{}}eFPGA\\ size\end{tabular}} &
  \multicolumn{3}{c}{\begin{tabular}[c]{@{}c@{}}eFPGA\\ params (N-K)\end{tabular}} &
  \multicolumn{3}{c}{CLB Util. [\%]} &
  \multicolumn{3}{c}{I/O Util. [\%]} \\ \cmidrule(l){3-5} \cmidrule(l){6-8} \cmidrule(l){9-11} \cmidrule(l){12-14}\cmidrule(l){15-17}    
 &
   &
  \multicolumn{1}{c}{ALICE} &
  \multicolumn{1}{c}{NK} &
  \multicolumn{1}{c}{KN} &
  \multicolumn{1}{c}{ALICE} &
  \multicolumn{1}{c}{NK} &
  \multicolumn{1}{c}{KN} &
  \multicolumn{1}{c}{ALICE} &
  \multicolumn{1}{c}{NK} &
  \multicolumn{1}{c}{KN} &
  \multicolumn{1}{c}{ALICE} &
  \multicolumn{1}{c}{NK} &
  \multicolumn{1}{c}{KN} &
  \multicolumn{1}{c}{ALICE} &
  \multicolumn{1}{c}{NK} &
  \multicolumn{1}{c}{KN} \\ \midrule
\multirow{9}{*}{\rotatebox[origin=c]{90}{\makecell{cfg1:\\64 I/O bits\\\& 2 eFPGAs}}} & DES3     
            & \makecell{sbox4, sbox3,\\sbox2;\\sbox8, sbox7,\\ sbox6, sbox5} & \makecell{sbox8, sbox3,\\sbox2, sbox1;\\sbox7, sbox6,\\sbox5, sbox4} & \makecell{sbox8, sbox3,\\sbox2, sbox1;\\sbox7, sbox6,\\sbox5, sbox4}
            & \makecell{6x6\\\\7x7} & \makecell{4x4\\\\4x4} &  \makecell{4x4\\\\4x4} & \makecell{4-4 \\\\ 4-4} & \makecell{4-6\\\\4-6} & \makecell{4-6\\\\4-6} & \makecell[r]{100\\\\100} & \makecell[r]{100\\\\100} & \makecell[r]{100\\\\100} & \makecell[r]{23\\\\25} & \makecell[r]{100\\\\62} & \makecell[r]{100\\\\62} \\
            & FIR      & \makecell{right mul.\\block} & \makecell{right mul.\\block} & \makecell{right mul.\\block} & 6x6 & 5x5 & 5x5 & 4-4 & 6-4 & 7-3 & 69 & 78 & 89 & 50 & 67 & 67  \\
            & SHA256   & \makecell{k constants} & \makecell{k constants} & \makecell{k constants} & \makecell{11x11} & \makecell{4x4} & \makecell{4x4} & 4-4 & 8-6 & 8-6 & 85 & 100 & 100 & 13 & 59 & 59 \\
            & SASC     & \makecell{sasc fifo} & \makecell{sasc fifo} & \makecell{sasc fifo} & \makecell{7x7} & \makecell{5x5} & \makecell{5x5} & 4-4 & 6-6 & 8-4 & 76 & 89 & 100 & 14 & 24 & 24 \\
            & USB\_PHY & \makecell{usb tx phy} & \makecell{usb tx phy} & \makecell{usb tx phy} & 7x7 & 5x5 & 5x5 & 4-4 & 7-6 & 8-5 & 80 & 100 & 89 & 11 & 18 & 18 \\
            & GCD      & \makecell{UnitCtr, Mux;\\RegEn} & \makecell{ZeroComp, RegEn;\\UnitCtr, Mux} & \makecell{ZeroComp, RegEn;\\UnitCtr, Mux} & \makecell{5x5\\4x4} &  \makecell{4x4\\4x4} & \makecell{4x4\\4x4} & \makecell{4-4\\4-4} &  \makecell{6-6\\8-5} & \makecell{8-3\\8-5} & \makecell[r]{100\\100} & \makecell[r]{100\\100} & \makecell[r]{100\\100} & \makecell[r]{65\\53} & \makecell[r]{97\\80} & \makecell[r]{97\\80} \\
            & NAUTILUS & \makecell{ALU, LIFO;\\RegFile} & \makecell{RegFile, ShiftByte;\\ALU} & \makecell{RegFile, ShiftByte;\\ALU} & \makecell{13x13\\9x9} & \makecell{6x6\\4x4} & \makecell{6x6\\4x4} & \makecell{4-4\\4-4} & \makecell{8-6\\8-7} & \makecell{8-6\\8-7} &  \makecell[r]{90\\86} &  \makecell[r]{94\\100} &  \makecell[r]{94\\100} &  \makecell[r]{16\\16} &  \makecell[r]{42\\45} &  \makecell[r]{42\\45}\\ 
         \midrule
\multirow{7}{*}{\rotatebox[origin=c]{90}{\makecell{cfg2:\\96 I/O bits\\\& 1 eFPGA}}}  & DES3     
            & \makecell{sbox8, sbox7,\\ sbox6, sbox5,\\ sbox4, sbox1} & \makecell{sbox8, sbox7,\\ sbox6, sbox5,\\ sbox4, sbox3,\\sbox2, sbox1} &  \makecell{sbox8, sbox7,\\ sbox6, sbox5,\\ sbox4, sbox3,\\sbox2, sbox1}
                                & 8x8 & 5x5 & 5x5 & 4-4  & 4-6 & 4-6 & 100 & 89 & 89 & 31 & 83 & 83 \\
            & FIR      & \makecell{block right} & \makecell{block right} & \makecell{block right} & 6x6 & 5x5 & 5x5 & 4-4 & 5-6 & 7-3 & 88 & 100 & 100 & 52 & 69 & 69 \\
            & SHA256   & \makecell{k constants} & \makecell{k constants} & \makecell{k constants} & \makecell{11x11} & \makecell{4x4} & \makecell{4x4} & 4-4 & 8-6 & 8-6 & 85 & 100 & 100 & 13 & 59 & 59 \\
            & SASC     & \makecell{sasc fifo} & \makecell{sasc fifo} & \makecell{sasc fifo} & \makecell{7x7} & \makecell{5x5} & \makecell{5x5} & 4-4 & 6-6 & 8-4 & 76 & 89 & 100 & 14 & 24 & 24\\
            & USB\_PHY & \makecell{usb tx phy} & \makecell{usb tx phy} & \makecell{usb tx phy} & 7x7 & 5x5 & 5x5 & 4-4 & 7-6 & 8-5 & 80 & 100 & 89 & 11 & 18 & 18 \\
            & GCD      & \makecell{ZeroComp, Mux} & \makecell{UnitCtr, Mux} & \makecell{UnitCtr, Mux} & 5x5 & 4x4 & 4x4 & 4-4 & 8-5 & 8-5 & 100 & 100 & 100 & 69 & 97 & 97 \\
            & NAUTILUS & \makecell{Control Unit} & \makecell{Control Unit} & \makecell{Control Unit} & 8x8 & 6x6 & 6x6 & 4-4 & 5-7 & 8-4 & 83 & 100 & 100 & 48 & 72 & 72 \\ 
           \bottomrule
\end{tabular}%
}
\end{table*}

\begin{figure*}[h]
    \hspace{-15pt}
    \centering
    \begin{tabular}{ccc}
        \begin{subfigure}{0.5\textwidth}
            \includegraphics[width=\linewidth]{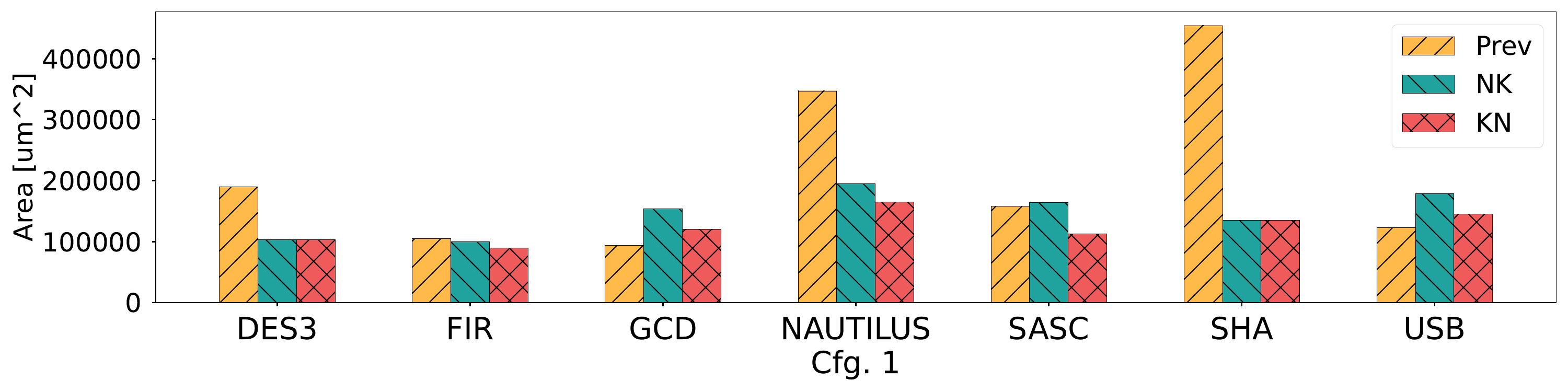}
            \vspace{-17pt}
        \end{subfigure} & \hspace{-10pt}
        \begin{subfigure}{0.5\textwidth}
            \includegraphics[width=\linewidth]{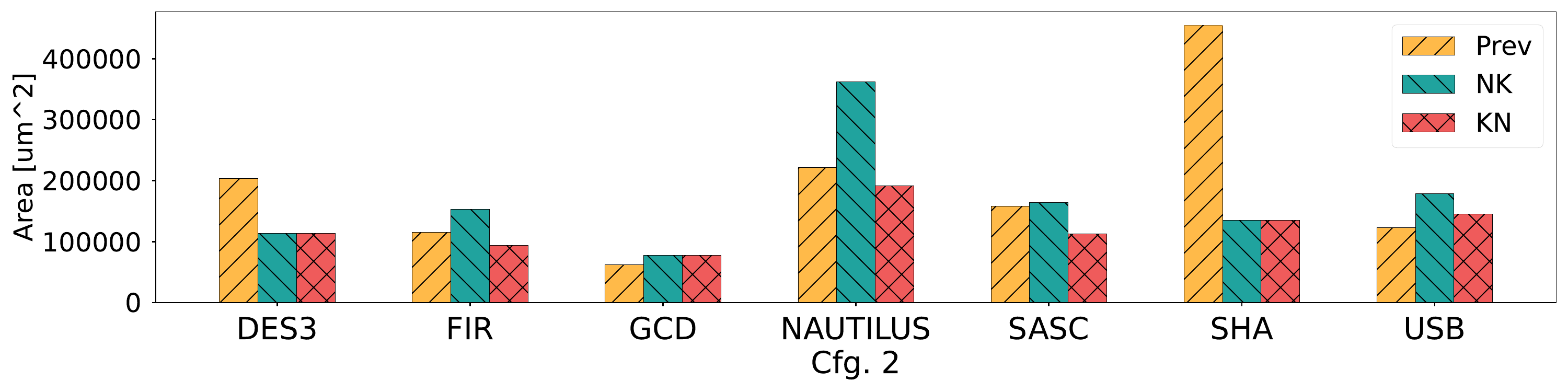}
            \vspace{-17pt}
        \end{subfigure} \\ 
        \begin{subfigure}{0.5\textwidth}
            \includegraphics[width=\linewidth]{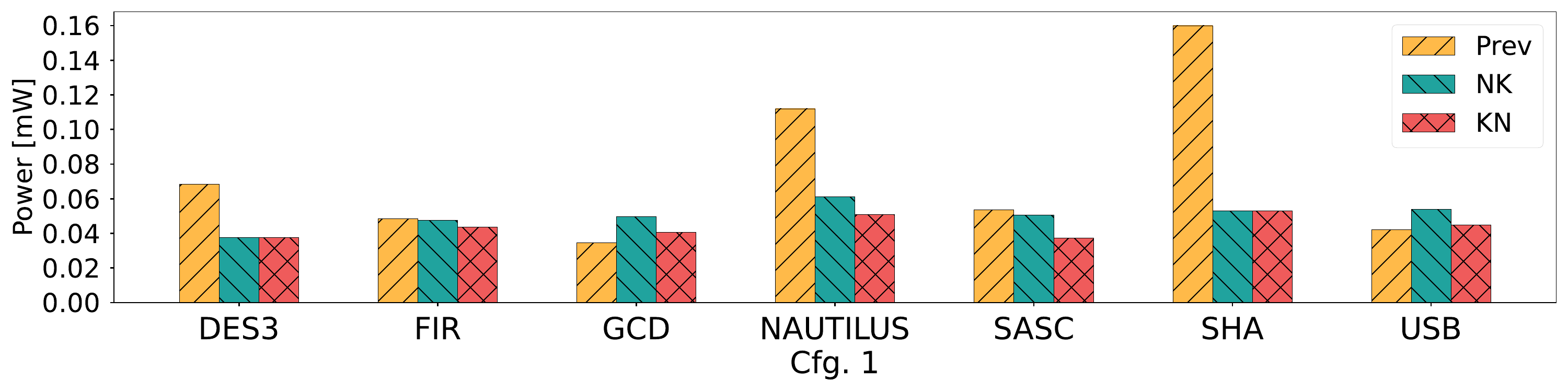}
            \vspace{-17pt}
        \end{subfigure} & \hspace{-10pt}
        \begin{subfigure}{0.5\textwidth}
            \includegraphics[width=\linewidth]{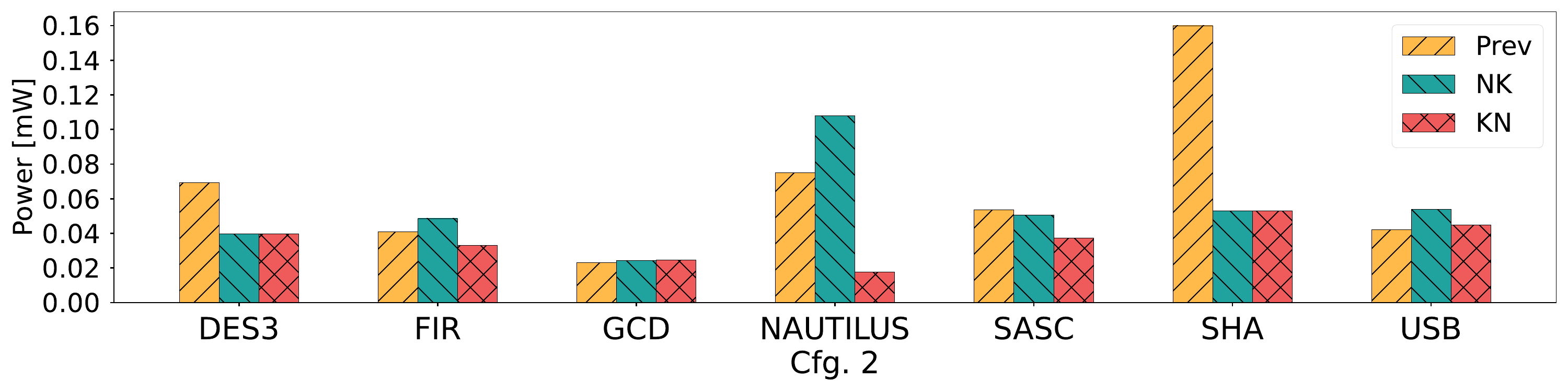}
            \vspace{-17pt}
        \end{subfigure} \\ 
        \begin{subfigure}{0.5\textwidth}
            \includegraphics[width=\linewidth]{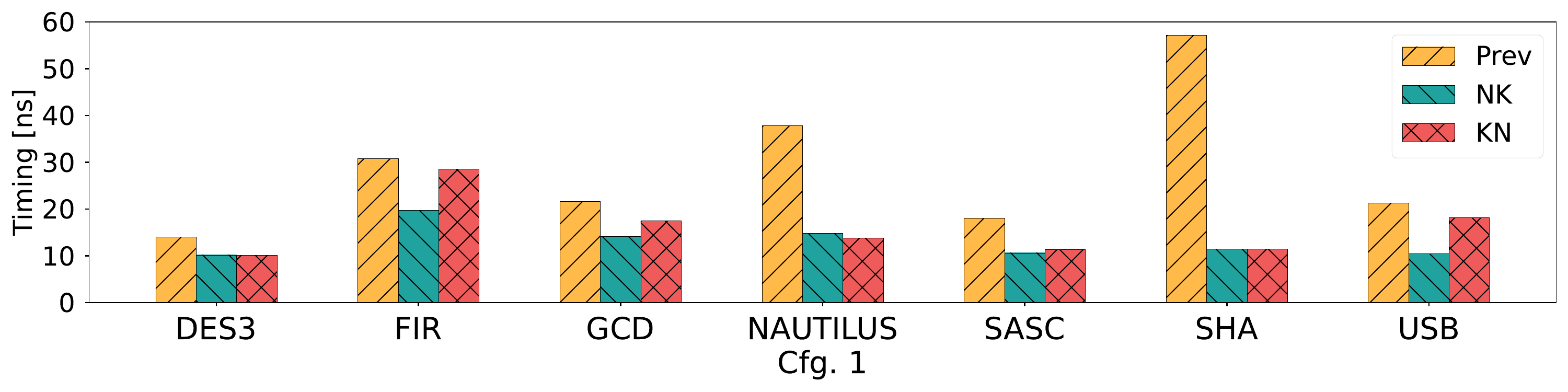}
            \vspace{-17pt}
        \end{subfigure} &\hspace{-10pt}
        \begin{subfigure}{0.5\textwidth}
            \includegraphics[width=\linewidth]{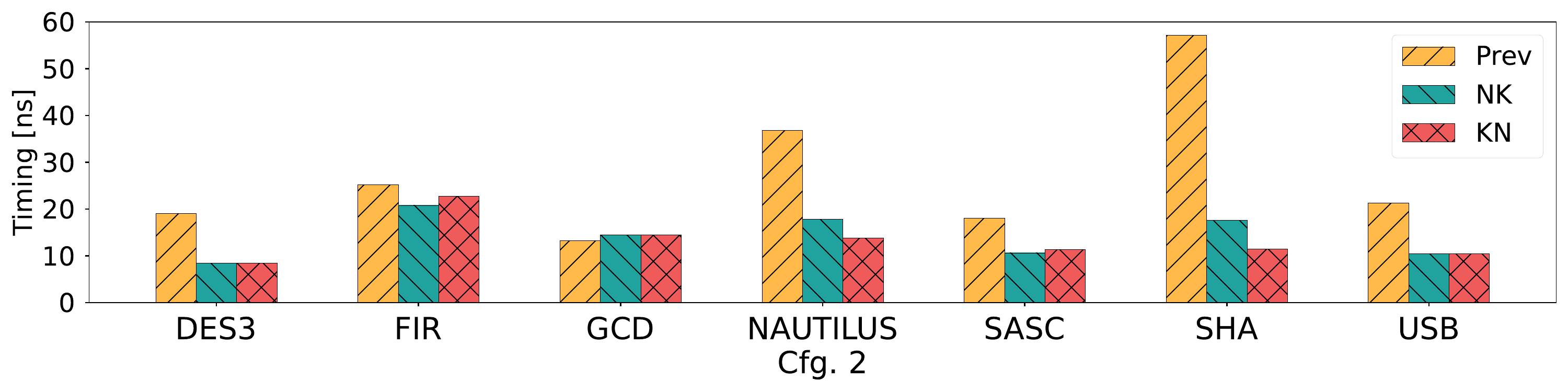}
            \vspace{-17pt}
        \end{subfigure}
    \end{tabular}
    \caption{Area, Power, and Timing results. The proposed heuristics outperform the state of the art, reducing overheads.}
    \label{tab:results3}
\end{figure*}

In~\autoref{tab:results}, we present the results at each flow step. In \autoref{tab:results2}, we present the final redacted solutions with their respective CLB and I/O utilization. In \autoref{tab:results3}, we present the overhead results.

Diving into \autoref{tab:results}, we can see how the module instances are filtered at the first phase, which identifies the candidate modules set $R$. This first phase also includes the dataflow analysis needed for the module filtering. The big fluctuations of the time needed in this phase across the different benchmarks reflect the dataflow complexity of the designs. After module filtering, combinations of these modules are identified in the cluster identification phase, which yields the candidate module cluster $C$. Here, we can see that the size of $C$ can drastically increase from the size of $R$ when the candidate modules are all independent (like the Sbox modules in DES3). 
The module filtering and cluster identification phases are performed in the same way for all flows. For this reason, we did not report separate data as we did for the fabric exploration phase. From the fabric exploration and eFPGA selection phase, we can see how the computational load is higher with the heuristics as they need to perform more OpenFPGA runs for the parameter exploration. The runtime is still reasonable for an EDA flow. From \autoref{tab:results2} and \autoref{tab:results3}, we can see that this computational overhead is well spent. We can identify different scenarios. For DES3, we can see how, in both configurations, the heuristics allow us to redact more modules while getting lower overheads (almost 2$\times$ lower). The redacted modules for FIR, SHA256, and SASC do not change, though tailoring the fabrics gets up to 3.3$\times$ lower overheads. For GCD and Nautilus, the redacted modules change for bigger ones, which increases the overheads by 1.2$\times$. USB\_PHY shows increased area overheads with improvements only in the timing for the heuristics while redacting the same module. This configuration finds a case where more but smaller tiles end up in a lower area than fewer bigger tiles. The latter configuration (identified by KN heuristics) reduces the timing overhead by $\sim$50\% while increasing area by $\sim$15\%. 
In DES3, using the heuristics, the final redacted modules are the same in \texttt{cfg1} and \texttt{cfg2}, with \texttt{cfg1} splitting the redacted modules in 2 4$\times$4 eFPGAs and \texttt{cfg2} using a single 5$\times$5 eFPGA. From the synthesis results, we can see that the former presents lower overheads, but from our security analysis in \autoref{sec:sec_identification} 4$\times$4 eFGPAs are orders of magnitude weaker to SAT attacks.   

In all cases, the I/O utilization is improved. DES3 is the only benchmark for which the CLB utilization does not improve, although the I/O utilization for DES3 has the biggest improvements across our benchmarks. Higher fabric utilization means we are wasting fewer resources and that the solutions are more resilient from attacks~\cite{mohan_hardware_2021,our_iccad_21,bhandari2021fabrics}.

\begin{figure}[t]
 \begin{subfigure}[b]{0.47\columnwidth}
 \centering
 \includegraphics[height=3.30cm]{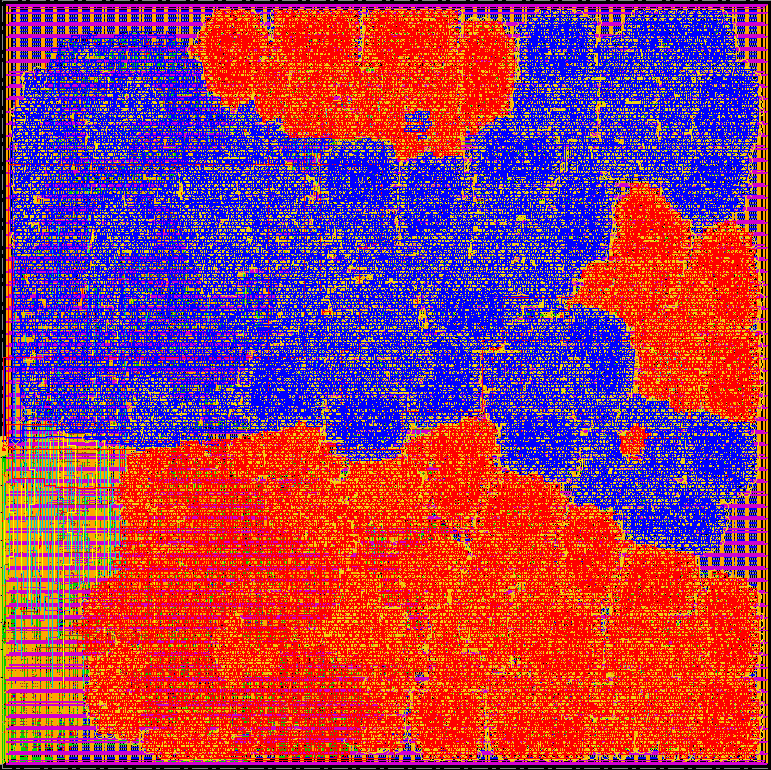}
 \caption{\texttt{cfg1}: two 4$\times$4 N=4; K=6 eFPGA (103,007$\mu m^2$)}\label{fig:gcd_2}
 \end{subfigure}
 \hfill
 \begin{subfigure}[b]{0.47\columnwidth}
 \centering
 \includegraphics[height=3.47cm]{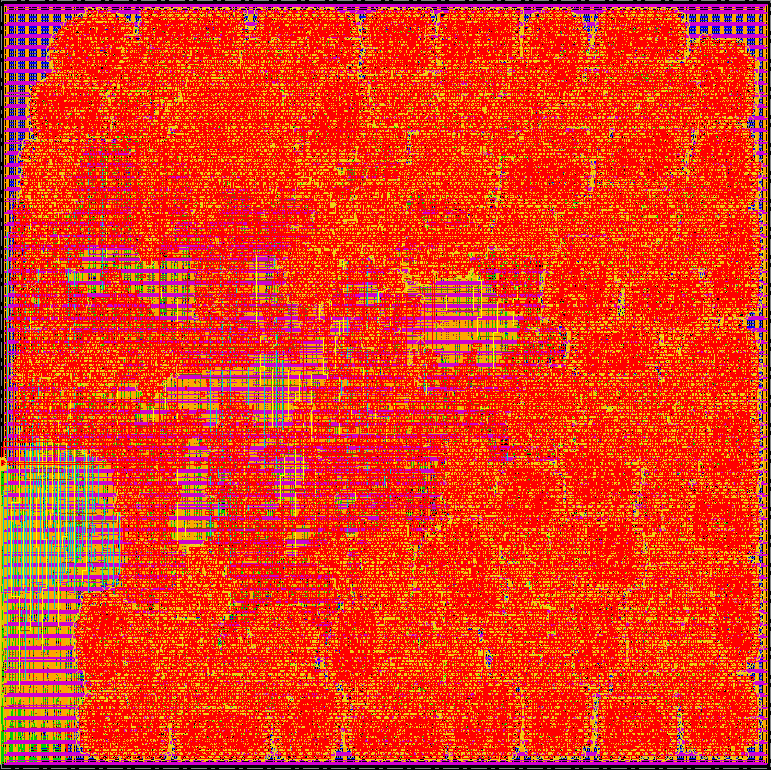} 	
 \caption{\texttt{cfg2}: one 5$\times$5 N=4; K=6 eFPGA (113,723$\mu m^2$)}\label{fig:gcd_1}
 \end{subfigure}
\vspace{-3pt}\caption{Physical layouts of DES3 solutions using the NK heuristic. In both configurations, the redacted modules are the same, on the left distributed between two eFPGAs and on the right all into the same eFPGA. The proportions are to scale.}\label{fig:layouts}	
\end{figure}

Figure~\ref{fig:layouts} shows the two physical designs for DES3 using the NK heuristic (which, by chance, coincides with the KN ones). This benchmark is small, so most of the chip is occupied by the eFPGA(s). However, the overhead will become less relevant when the component is inserted into a larger system-on-chip (like PicoSoc in reference~\cite{our_iccad_21}). We can notice how, despite redacting the same modules, using two smaller eFPGAs yields a slightly smaller area ($\approx$10\%). A designer might prefer this solution as it would require an attacker to retrieve 2 bitstreams instead of one.

Looking at the two proposed heuristics, both yield bespoke fabrics that improve redaction solutions. The KN heuristics often yield better synthesis results with lower area and power overheads. A designer might want to use the KN heuristic first and then use NK only if the results are unsatisfactory. Using both heuristics is still much faster than the exhaustive approach (see \autoref{tab:heu}).

Comparing our results with SheLL~\cite{shell} is not straightforward as the set of benchmarks differs considerably, and the only common benchmark is FIR. They claim up to ~2$\times$ reduction in overheads compared to the ALICE~\cite{alice} approach, whereas our results show up to a 3.3$\times$ reduction (for SHA). 

In cases where the final eFPGA size and fabric configuration are not considered secure from the secure fabric identification step, the designer has two options: discard the solution or integrate additional security measures to mitigate the attacks. For example, in our case, 4$\times$4 fabrics are not secure against the IcySAT attack that we considered for our secure fabric identification. The designer could still choose 4$\times$4 fabrics if they also integrate a secure scan chain protection technique like DisORC~\cite{disorc}.

\section{Conclusions and Future Work} \label{sec:conclusion}

This paper proposes ARIANNA, an expanded version of the ALICE framework, focusing on the eFPGA fabric parameter selection problem. We proposed two heuristics for the design space exploration of fabric parameters for eFPGA redaction. We first analyzed the heuristics isolated in the parameter selection phase against an exhaustive approach and then in the complete framework against the previous work. 

Our results showed significant improvements (up to 3.3$\times$ lower overheads and 4$\times$ higher fabric utilization) can be obtained by tailoring bespoke fabrics for eFPGA redaction. Compared to state-of-the-art, we find that SheLL~\cite{shell} reduced overheads by 55\% compared to ALICE, whereas our work reached improvements of up to 330\%.

Our heuristics allow for a more scalable framework with respect to an exhaustive approach. Moreover, the proposed heuristics yield higher utilization of the eFPGA fabric, meaning less wasted resources and better security~\cite{mohan_hardware_2021,our_iccad_21,bhandari2021fabrics}.
With this contribution, ARIANNA is now a complete framework that can tackle both the module selection and the fabric configuration problems faced when applying eFPGA redaction. Future work includes expanding the framework to explore more eFPGA configuration parameters for the fabrics and the overall architecture. This step can include ML-driven exploration techniques. Also, selection methods can be extended to include more criteria or perform a fine-grain decomposition and redaction of larger modules.





\bibliographystyle{IEEEtran}
\bibliography{main.bib}

\end{document}